\documentclass[12pt,fleqn]{article}
\usepackage{amssymb,amsmath}

- 1.5cm
\renewcommand{\theequation}{\arabic{section}.\arabic{equation}}
\renewcommand{\thesection}{\arabic{section}.}
\renewcommand{\thesubsection}{\arabic{section}.\arabic{subsection}.}
\catcode`@=11 \@addtoreset{equation}{section} \catcode`@=12

\newcommand{\SGamma}{\varGamma}
\newcommand{\Tr}{{\mathrm{Tr\,}}}
\begin{document}

\title{\textbf{Nonperturbative heat kernel and nonlocal effective
action}\footnote{Based on the talk given at the Workshop "Gravity
in Two Dimensions", Erwin Schroedinger Institute for Mathematical
Physics, Vienna, October 2003.}}
\date{}
\author{A.O.Barvinsky
and D.V.Nesterov\\
\emph{Theory Department, Lebedev Physics Institute,}\\
\emph{Leninsky prospect 53, Moscow 119991, Russia}}

\maketitle

\begin{abstract}
We present an overview of recent nonperturbative results in the
theory of heat kernel and its late time asymptotics responsible
for the infrared behavior of quantum effective action for massless
theories. In particular, we derive the generalization of the
Coleman-Weinberg potential to physical situations when the field
is not homogeneous throughout the whole spacetime. This
generalization represents a new nonlocal and nonperturbative
action accounting for the effects of a transition domain between
the spacetime interior and its infinity. In four dimensions these
effects delocalize the logarithmic Coleman-Weinberg potential,
while in $d>4$ they are dominated by new powerlike and
renormalization-independent nonlocal structure. Nonperturbative
behavior of the heat kernel is also constructed in curved
spacetime with asymptotically-flat geometry, and its conformal
properties are analyzed for conformally invariant scalar field.
The problem of disentangling the local cosmological term from
nonlocal effective action is discussed.
\end{abstract}

%PACS: {04.62.+v, 11.10.Lm, 11.15.Tk}

%Keywords:
%\textit{nonlocal heat kernel expansion, effective action}

\newpage
\tableofcontents
\newpage

\section{Introduction: heat kernel and effective action}
\hspace{\parindent} It is widely recognized that nonlocal
phenomena play a very important role in quantum physics. In
contrast to low-energy vacuum polarization effects by massive
quantum fields they characterize high-energy asymptotics in
massive theories or infrared behavior in theories of massless
fields. Even in models with well-established low-energy behavior
like Einstein gravity these phenomena become increasingly
interesting due to the attempts of resolving the cosmological
constant and acceleration problems by means of nonlocal
long-distance modifications of the theory. These modifications
often call for nonperturbative treatment in view of the
nonperturbative aspects of van Damm-Veltman-Zakharov discontinuity
problem \cite{VVDZ} and the presence of a hidden nonperturbative
scale in gravitational models with extra dimensions \cite{scale}.

On the other hand, nonlocalities also (and, moreover, primarily)
arise in virtue of fundamental quantum  effects of matter and
graviton loops which, for instance, can play important role in
gravitational radiation theory \cite{MWZ,vilkov} and cosmology
\cite{Woodard}. Therefore, they can successfully compete with
popular phenomenological mechanisms of infrared modifications,
induced, say, by braneworld scenarios with extra dimensions
\cite{GRS,DGP} or other models \cite{WoodMOND}. This makes
nonperturbative analysis of nonlocal quantum effects very
interesting and promising.

A basic tool for the description of these effects is the quantum
effective action and the heat kernel technique of its calculation.
For a generic theory of the field $\varphi(x)$ the classical
action $S[\varphi]$ gives rise to the inverse propagator -- the
operator of linear field disturbances on the background of
$\varphi(x)$
    \begin{eqnarray}
     S[\varphi]\;\rightarrow \;F(\nabla)\delta(x,y)
     =\frac{\delta^2 S[\varphi]}
     {\delta\varphi(x)\delta\varphi(y)}.  \label{1.1}
    \end{eqnarray}
Then the quantum effective action $\SGamma[\varphi]$ follows from
its classical counterpart as a loop expansion in powers of $\hbar$
    \begin{equation}
     S[\varphi]\rightarrow\SGamma
     =S[\varphi]+\hbar\SGamma_{1-loop}[\varphi]
     +\hbar^2\SGamma_{2-loop}[\varphi]+...        \label{1.2}
    \end{equation}
The first few orders of this expansion are graphically depicted
below as Feynman graphs with the propagator -- the Green's
function of the operator (\ref{1.1}) -- and relevant vertices
calculated on the background of a generic $\varphi$
    \begin{eqnarray}
     \SGamma_{1-loop}=\frac12 {\mathrm{Tr}} \,\ln F(\nabla)
     =\frac12
       \begin{picture}(0,0)
        \put(20,3){\circle{28}}
       \end{picture}                      \label{1.3}
    \end{eqnarray}

    \begin{equation}
     \SGamma_{2-loop}=\frac18
       \begin{picture}(0,0)
        \put(15,13){\circle{18}}
        \put(15,3){\circle*{4}}
        \put(15,-7){\circle{18}}
       \end{picture}
     \qquad\quad+\frac1{12}\,
       \begin{picture}(0,0)
        \put(15,3){\circle{28}}
        \put(15,17){\circle*{4}}
        \put(15,-11){\circle*{4}}
        \put(15,-11){\line(0,28){28}}
       \end{picture}
       \qquad\quad.\label{1.4}
    \end{equation}
The one-loop part (\ref{1.3}) is peculiar in that it does not
explicitly contain the vertices of the classical action (unless it
is expanded in powers of the mean field $\varphi$) and given by
the functional trace of the logarithm of $F(\nabla)$.

In local field theories without loss of generality the operator
(\ref{1.1}) has the form
    \begin{equation}
     F(\nabla)=\Box-V(x),                   \label{1.5}
    \end{equation}
where $V(x)$ is some potential and
    \begin{equation}
     \Box=g^{\mu\nu}\nabla_\mu\nabla_\nu      \label{1.6}
    \end{equation}
is a covariant d'Alembertian in curved spacetime with metric
$g_{\mu\nu}$.

A very efficient way of analyzing Feynman graphs of the form
(\ref{1.3})-(\ref{1.4}) on {\it generic field background}, i. e.
with generic metric and potential, is based on the use of the heat
kernel
    \begin{equation}
     K(s|\,x,y)=\exp(sF(\nabla))\,\delta(x,y).   \label{1.7}
    \end{equation}
It solves the heat equation with unit initial condition at $s=0$
    \begin{equation}
     \frac{\partial K(s)}{\partial s}
     =F(\nabla)\,K(s),\quad K(0)=\mathbb{I}   \label{1.8}
    \end{equation}
and generates as a result of integration over the proper time
parameter the main ingredient of the Feynman diagrammatic
technique -- the propagator of the operator (\ref{1.1})
    \begin{eqnarray}
     G(x,y)\equiv\frac1{F(\nabla)}\,\delta(x,y)
     =-\int_{0}^{\infty }\!ds\:K(s|\,x,y)      \label{1.9}
    \end{eqnarray}
and, in closed form, the one-loop effective action
    \begin{eqnarray}
     &&\SGamma_{1-loop}=\frac{1}{2}\,
     \int_{0}^{\infty }\frac{ds}{s}\:
     \Tr K(s)\,,                                \label{1.10}\\
     &&\Tr K(s)=\int dx\, K(s|\,x,x).       \label{1.11}
    \end{eqnarray}

The efficiency of the heat kernel and proper time method are based
on the well-known and universal behavior of $K(s|x,y)$ for a
generic second-order operator (\ref{1.5}) at $s\to 0$. This limit
is responsible for ultraviolet divergences and anomalies in field
theory, renormalization and low-derivative expansion underlying
vacuum polarization effects.

On the contrary, nonlocal terms arise as a contribution of the
upper limit in the proper-time integral (\ref{1.10}), which makes
the late time asymptotics of ${\Tr}K(s)$ most important for
another class of effects including particle creation and
scattering\footnote{Here the effective action is defined in
Euclidean space with positive-signature metric. Its application in
physical spacetime with Lorentzian signature is based on analytic
continuation methods which range from a conventional Wick rotation
in scattering theory (for in-out matrix elements) to a special
retardation prescription in a wide class of problems for a mean
field (in-in expectation value) \cite{CPTI,nonlocal}. These
methods nontrivially apply to nonlocal terms and extend from the
usual perturbation theory to its partial resummation corresponding
to the nonperturbative technique of the present work.}. Its
integrand -- the heat kernel trace, including its late time
asymptotics, was first studied within the covariant nonlocal
curvature expansion in \cite{CPTI,CPTII,CPTIII,asymp}. The goal of
this paper is to give a critical overview of these old results and
present the latest development in heat kernel theory concerning
its nonperturbative asymptotics in flat spacetime \cite{nnea}, its
generalization to curved spacetime geometry \cite{nneag} and
application in the theory of nonlocal effective action.

\section{Approximation schemes and infrared behavior}
\hspace{\parindent} To simplify the presentation, throughout this
section we will work in flat spacetime with metric
$g_{\mu\nu}=\delta_{\mu\nu}$ and, where necessary, briefly mention
relevant modifications due to spacetime curvature. Also we will
consider the case of a scalar field without spin, when the heat
kernel (\ref{1.7}) is a biscalar object without spin labels. In
this case the small-time behavior of the heat kernel, which
actually accounts for the success of renormalization scheme in
local quantum  field theory, looks especially simple and reads
    \begin{eqnarray}
     && K(s|\,x,y)=\frac1{(4\pi s)^{d/2}}\,
     \exp{\left\{-\frac{|x-y|^2}{4s}\right\}}\;
     \Omega(s|\,x,y)                                  \label{2.1}\\
     && \Omega(s|\,x,y)\rightarrow 1,\quad s\rightarrow 0,
    \end{eqnarray}
where $d$ is the spacetime dimensionality. This semiclassical
ansatz for the heat kernel guarantees that at $s\to 0$ it tends to
the delta-function $\delta(x,y)$ and contains all nontrivial
information about the potential $V(x)$ in the function
$\Omega(s|\,x,y)$ which is analytic at $s=0$. Its expansion in
powers of $s$ underlies the technique of local Schwinger-DeWitt
expansion which looks as follows.

\subsection{Schwinger-DeWitt technique of local expansion}
\hspace{\parindent}From the heat equation (\ref{1.8}) one easily
derives a set of recurrent equations for the coefficients of
small-time expansion of $\Omega(s|\,x,y)$
    \begin{equation}
     \Omega (s|\,x,y)=\sum\limits_{n=0}^{\infty}
     a_{n}(x,y)\,s^{n},\quad s\rightarrow 0.  \label{2.2}
    \end{equation}
These coefficients play a very important role in quantum field
theory and have the name of HAMIDEW coefficients that was coined
by G.Gibbons to praize joint efforts of mathematicians and
physicists in heat kernel theory and its implications. The
equations for $a_{n}(x,y)$ can in closed form be solved for their
coincidence limits $a_{n}(x,x)$ in terms of the potential $V(x)$
and its derivatives. For the operator (\ref{1.5})-(\ref{1.6}) the
first few of them read
    \begin{eqnarray}
     && a_0(x,x)=1
     \nonumber\\
     && a_1(x,x)=-V(x)+\frac16 R(x)
     \nonumber\\
     && a_2(x,x)=\frac12\, V^2(x)+\frac16\, \Box V(x)
     -\frac16\, R(x) V(x)+
     O(R^{\;2}_{\mu\nu\alpha\beta}),     \label{2.3}
    \end{eqnarray}
where we also included the contribution of spacetime curvature
scalar $R$ and symbolically denoted the quadratic contribution of
curvature tensor in the second-order coefficient. As is clearly
seen, these quantities are local functions built of the
coefficients of the original differential operator and their
derivatives. The dimensionality of $a_{n}(x,x)$ in units of
inverse length grows with $n$ and is comprised of the powers of
dimensionful quantities $V(x)$, $R_{\mu\nu\alpha\beta}$ and their
derivatives.

Suppose now that instead of the theory with the operator
(\ref{1.5}) we consider the theory of massive field with a large
mass $m$. This corresponds to the replacement of the original
operator by
    \begin{equation}
     F(\nabla)\rightarrow F(\nabla)-m^2.    \label{2.4}
    \end{equation}
The corresponding heat kernel (\ref{2.1}) under this replacement
obviously acquires the overall exponential factor $\exp(-sm^2)$
damping the contribution of large values of $s$ in the proper time
integral (\ref{1.10})
    \begin{eqnarray}
     K(s|\,x,y)=\frac1{(4\pi s)^{d/2}}\,
     \exp{\left\{-\frac{|x-y|^2}{4s}-sm^2\right\}}\;
     \Omega(s|\,x,y).                                  \label{2.5}
    \end{eqnarray}
Substituting it together with the expansion (\ref{2.2}) into
(\ref{1.10}) and integrating the resulting series term by term we
obtain the one-loop effective action of massive theory in the form
of the asymptotic $1/m^2$ expansion
\cite{CPTII,nnea,PhysRep,DeWitt}
    \begin{eqnarray}
     &&\frac12\,{\Tr}\ln\left[F(\nabla)-m^2\right]
     =-\frac{1}{2\left( 4\pi \right) ^{d/2}}\int
     dx \int_{0}^{\infty}
     \frac{ds}{s^{d/2+1}}\:e^{-sm^{2}}
     \sum\limits_{n=0}^{\infty }\,s^n\,a_{n}(x,x)
     \nonumber\\
     &&\qquad=\SGamma_{\mathrm{div}}
     +\SGamma_{\log }-\frac{1}{2}
     \left( \frac{m^{2}}{4\pi}\right)^{d/2}
     \int dx\,\sum\limits_{n=d/2+1}^{\infty} \,
     \frac{\Gamma(n\!-\!d/2)}{(m^{2})^{n}}\,a_{n}(x,x).  \label{2.6}
    \end{eqnarray}
The first $d/2$ integrals (we assume that $d$ is even) are
divergent at the lower limit and generate ultraviolet divergences
$\SGamma_{\mathrm{div}}$ given by the first $d/2$ Schwinger-DeWitt
coefficients. In dimensional regularization they read
    \begin{eqnarray}
    &&\SGamma_\mathrm{div}\!=\!\frac{1}{2(4\pi )^{d/2}}\int
    dx\,\sum\limits_{n=0}^{d/2}
    \left[\frac{1}{\omega -d/2}
    -\Gamma ^{\prime }\Big(
    \frac{d}{2}-n+1\Big)\right]
    %\nonumber\\
    %&&\qquad\qquad\qquad\qquad\qquad\qquad\qquad\qquad
    %\times
    \frac{(-m^2)^{d/2-n}}{(d/2
    -n)!}\,a_{n}(x,x).       \label{2.6a}
    \end{eqnarray}
The logarithmic divergences are also accompanied by the
logarithmic term
    \begin{equation}
    \SGamma_{\log }=\frac{1}{2(4\pi )^{d/2}}\int
    dx\,\sum\limits_{n=0}^{d/2}\,
    \frac{(-m^{2})^{{d/2}-n}}{\left({d/2}
    -n\right) !}
    \ln \frac{m^{2}}{\mu ^{2}}\,a_{n}(x,x),       \label{2.6b}
    \end{equation}
containing the renormalization mass parameter $\mu^2$ reflecting
the renormalization ambiguity.

In the present form each term in the finite part of the action
(\ref{2.6}) is local, but this locality holds only in the range of
applicability of this expansion when the mass  is large and the
terms of the asymptotic series rapidly decrease with the growth of
$n$. This occurs when the mass parameter $m$ is large compared
with \emph{both} the derivatives of the potential and the
potential itself
    \begin{eqnarray}
     1\gg \frac{a_n}{(m^2)^n}
     \sim\left(\frac{V}{m^2}\right)^n\!,\;
     \left(\frac{\nabla}{m}\right)^{\!k}
     \!\!\!\left(\frac{V}{m^2}\right)^{\!l}\,,\quad
     k+2l=2n\,.  \label{2.7a}
    \end{eqnarray}
In the presence of the gravitational field these restrictions
include also the smallness of spacetime curvature and its
covariant derivatives compared to the mass parameter. Thus the
local Scwinger-DeWitt expansion is applicable only for slow
varying fields of small amplitude compared to the mass scale of
the model.

For large field strengths or rapidly varying fields the
Schwinger-DeWitt expansion becomes inapplicable and completely
blows up in the massless limit $m\rightarrow 0$. So the question
arises, what is the structure of the effective action in this
situation and how to calculate it. Below we consider two
perturbation methods which improve the Schwinger-DeWitt technique
to extend it to the class of massless models and then, in Sect.3,
go over to the nonperturbative technique based on the
nonperturbative late-time asymptotics of the heat kernel.

\subsection{Modified Schwinger-DeWitt expansion}
\hspace{\parindent} The first method may be called the modified
Schwinger-DeWitt expansion, because it is based on the resummation
of this expansion originating from the replacement of the mass
term by the potential $V(x)$ of the operator (\ref{1.7}),
$m^2\rightarrow V(x)$. When the potential is positive-definite
(which we shall assume here) it can play the role of the damping
factor similar to the mass term in the integral (\ref{2.6}). For
this purpose we have to extract the exponential dependence on
$sV(x)$ from the function $\Omega(s|\,x,y)$,
$\Omega(s|\,x,y)=e^{-sV(x)}\tilde{\Omega}(s|\,x,y)$, and write
instead of (\ref{2.5}) the ansatz
    \begin{eqnarray}
     K(s|\,x,y)=\frac1{(4\pi s)^{d/2}}\,
     \exp{\left\{-\frac{|x-y|^2}{4s}-V(x)\right\}}\;
     \tilde{\Omega}(s|\,x,y),                            \label{2.8}
    \end{eqnarray}
where the new function has the expansion in $s$ with the modified
Schwinger-DeWitt coefficients
    \begin{equation}
     \tilde{\Omega} (s|\,x,y)
     =\sum\limits_{n=0}^{\infty }
     \tilde{a}_{n}(x,y)\,s^{n}.           \label{2.9}
    \end{equation}
Obviously this expansion represents the partial resummation of the
initial series (\ref{2.2}) in  powers of the undifferentiated
potential. New coefficients $\tilde{a}_{n}(x,x)$ in contrast to
old ones have fewer number of terms and do not contain
undifferentiated potential (and, therefore, except for
$\tilde{a}_0=1$ vanish in the absence of gravity for $\nabla
V=0$). The original coefficients can be expressed in terms of them
as finite-order polynomials in $V$ with coefficients built of the
gradients of potential
    \begin{equation}
    a_m(x,x)=\frac{(-V)^m}{m!}
    +\sum\limits_{n=1}^{d/2}
    \frac{(-V)^{m-n}}{(m-n)!}\;
    \tilde{a}_{n}(x,x)=\frac{(-V)^m}{m!}+O(\nabla V). \label{rel}
    \end{equation}

Now the proper time integral in (\ref{2.6}) even for $m^2=0$ has
an infrared cutoff  at $s\sim 1/V(x)$ and in this case the
effective action is similar to (\ref{2.6}), where $m^{2}$ is
replaced by $V(x)$ and $a_{n}(x,x)$ by $\tilde{a}_{n}(x,x)$
    \begin{equation}
     \sum\limits_{n}\frac{a_{n}(x,x)}{(m^{2})^{n}}\,
     \rightarrow
     \sum\limits_{n}
     \frac{\tilde{a}_{n}(x,x)}{V^{n}(x)}.   \label{2.10}
    \end{equation}
In particular, the ultraviolet divergences are given by the
massless limit of $\SGamma_\mathrm{div}$ and the logarithmic part
gives rise to the ($d$-dimensional) Coleman-Weinberg term
$\SGamma_{\log}\rightarrow\SGamma_\mathrm{CW}+O(\nabla V)$,
    \begin{eqnarray}
     \SGamma_\mathrm{CW}=\frac{1}{2(4\pi )^{d/2}}
     \int dx\, \frac{(-V)^{d/2}(x)}{(d/2)!}
     \ln \frac{V(x)}{\mu ^{2}},                 \label{2.11}
    \end{eqnarray}
corrected by the contribution due to the derivatives of $V(x)$
(confer Eq.(\ref{rel}) with $m=d/2$). $\SGamma_\mathrm{CW}$ here
is the spacetime integral of the Coleman-Weinberg effective
potential. For instance, in four dimensions in the
$\varphi^4$-model of the self-interacting scalar field with
$V(\varphi)\sim\varphi^2$, this is the original Coleman-Weinberg
effective potential, $\varphi^4\ln (\varphi^2/\mu^{2})/64\pi
^{2}$.

From (\ref{2.10}) it follows that, in contrast to the original
Schwinger-DeWitt series, its modified version represents the
expansion in the derivatives of $V$ rather than powers of $V$
itself. Indeed, the modified Schwinger-DeWitt coefficients do not
contain the undifferentiated potential and the typical structure
of the terms entering $\tilde{a}_{n}(x,x)$ is represented by $m$
derivatives acting in different ways on the product of $j$
potentials, $\nabla ^{m}V^{j}(x)$, where $m+2j=2n$. Every $V$ here
should be differentiated at least once and therefore $m\geq j$.
Thus the typical terms of the expansion (\ref{2.10}) can be
symbolically written down as
    \begin{equation}
    \frac{\tilde{a}_{n}}{V^n}\sim
    \sum\limits_{j=1}^{[2n/3]}
    \frac{\nabla ^{2n-2j}V^{j}}{V^n},                \label{2.12}
    \end{equation}
where the upper value of $j$ is the integer part of $2n/3$.
Therefore this expansion is efficient as long as the potential is
slowly varying in units of the potential itself
    \begin{equation}
    \frac{\nabla ^{2}V(x)}{V^{2}(x)}\ll 1,\quad
    \frac{\left( \nabla V(x)\right) ^{2}}
    {V^{3}(x)}\ll 1,\quad ...\, .                   \label{parexp}
    \end{equation}

When the potential is bounded from below by a large positive
constant this condition can be easily satisfied throughout the
whole spacetime. But this case is uninteresting because it
reproduces the original Schwinger-DeWitt expansion with $m^2$
playing the role of this bound. More interesting is the case of
the asymptotically empty spacetime when the potential and its
derivatives fall off to zero by the power law
    \begin{equation}
    V(x)\sim \frac{1}{|x|^{p}},\quad
    \nabla ^{m}V(x)\sim \frac{1}{|x|^{p+m}},
    \quad |x|\rightarrow \infty          \label{potfall}
    \end{equation}
for some positive $p$. For such a potential terms of the
perturbation series (\ref{2.10}) behave as
    \begin{equation}
    \frac{\tilde{a}_{n}(x,x)}{V^{n}(x)}\sim
    \sum\limits_{j=1}^{[2n/3]}|x|^{(p-2)(n-j)}   \label{exppar1}
    \end{equation}
and thus decrease with $n$ only if $p<2$. For $p\geq 2\,$ the
modified gradient expansion completely breaks down. It makes sense
only for slowly decreasing potentials of the form (\ref{potfall})
with $p<2$. In this case the potential $V(x)$ is not integrable
over the whole spacetime $\left( \int dx\,V(x)=\infty \right)$ and
moreover even the operation $(1/\Box )V(x)$ is not well
defined\footnote{For the convergence of the integral in
$\left(1/\Box \right) V$ the potential $V(x)$ should fall off at
least as $1/|x|^{3}$ in any spacetime dimension \cite{CPTII}.}.
Therefore, the above restriction is too strong to account for
interesting physical problems in which the parameter $p$ typically
coincides with the spacetime dimensionality $d$. In addition, the
modified asymptotic expansion is completely local and does not
allow one to capture nonlocal terms of effective action.

Thus an alternative technique is needed to obtain the late-time
contribution to the proper-time integral and, in particular, to
understand whether and when this integral exists at all in
massless theories. The answer to this question lies in the
late-time asymptotics of the heat kernel at $s\rightarrow\infty$
which can be perturbatively analyzed within the covariant
perturbation theory of \cite{CPTI,CPTII,CPTIII,asymp,basis}.

\subsection{Covariant perturbation theory}
\hspace{\parindent} In the covariant perturbation theory the full
potential $V(x)$ is treated as a perturbation and the solution of
the heat equation is found as a series in its powers. From the
viewpoint of the Schwinger-DeWitt expansion it corresponds to an
infinite resummation of all terms with a given power of the
potential and arbitrary number of derivatives. The result reads as
    \begin{equation}
    {\Tr}K(s)\equiv \int dx\,K(s|\,x,x)=
    \sum\limits_{n=0}^{\infty }{\Tr}K_{n}(s),      \label{CPT1}
    \end{equation}
where
    \begin{equation}
    {\Tr}K_{n}(s)=\int
    dx_{1}dx_{2}...dx_{n}\,
    F_{n}(s|\,x_{1},x_{2},...x_{n})
    \,V(x_{1})V(x_{2})...V(x_{n}),  \label{Kn}
    \end{equation}
and the nonlocal form factors $F_{n}(s|\,x_{1},x_{2},...x_{n})$
were explicitly obtained in \cite{CPTI,CPTII,CPTIII} up to $n=3$
inclusive. In the presence of gauge fields and gravity this
expansion can be easily generalized by including the fibre bundle
$\mathcal{R}_{\mu\nu}$ and Ricci curvature\footnote{In
asymptotically flat spacetime with natural vacuum boundary
conditions the Riemannian curvature can be perturbatively
expressed in terms of the Ricci tensor \cite{CPTIII,basis}, that
is why it does not enter the expansion as an independent entity.}
in the full set of perturbatively treated field strengths,
$V\rightarrow \mathcal{R}=(V,\mathcal{R}_{\mu\nu},R_{\mu\nu})$ and
covariantizing the corresponding nonlocal form factors.

It was shown \cite{CPTII} that at $s\rightarrow \infty $ the terms
in this expansion behave as
    \begin{equation}
    {\Tr}K_{n}(s)=
    O\left( \frac{1}{s^{d/2-1}}\right), \quad
    n\geq 1,                                \label{CPTas}
    \end{equation}
and, therefore in spacetime dimension $d\geq 3$ the integral in
(\ref{1.10}) is infrared convergent
    \begin{equation}
    \SGamma\sim\int\limits^{\infty }\frac{ds}{s}\,
    O\left( \frac{1}{s^{d/2-1}}\right)
    <\infty .                 \label{infCPT}
    \end{equation}
In one and two dimensions this expansion for $\SGamma$ does not
exist except for the special case of the massless theory in curved
two-dimensional spacetime, when it reproduces the Polyakov action
\cite{CPTII,FrolVilk,Polyakov}, which alternatively can be
obtained by integrating the conformal anomaly
\cite{FrolVilk,Polyakov}.
    \begin{equation}
     \SGamma_\mathrm{Polyakov}\sim\int d^2 x\,g^{1/2}\,R\frac1\Box R.
    \end{equation}

Covariant perturbation theory should always be applicable whenever
$d\geq 3$ and the potential $V$ is sufficiently small, so that its
effective action explicitly features analyticity in the potential
at $V=0$. Therefore, its serious disadvantage is that this theory
does not allow one to overstep the limits of perturbation scheme
and, in particular, discover non-analytic structures in the action
if they exist.

\section{Nonperturbative late-time asymptotics}
\hspace{\parindent} Nonperturbative technique for the heat kernel
is based on the approximation qualitatively different from those
of the previous section. Rather than imposing certain smallness
restrictions on the background fields we consider them rather
generic, but consider the limit of large proper time $s\to\infty$.
Continuing working in flat spacetime with
$g_{\mu\nu}=\delta_{\mu\nu}$, we substitute the ansatz (\ref{2.1})
in the heat equation and immediately obtain the following equation
for the unknown function $\Omega(s|\,x,y)$
    \begin{eqnarray}
    \frac{\partial\Omega}{\partial s}
    +\frac1s\,(x-y)^\mu\nabla_\mu\Omega
    =F(\nabla)\,\Omega.                     \label{3.1}
    \end{eqnarray}
Then we assume the existence of the following $1/s$-expansion for
this function (which follows, in particular, from the perturbation
theory for $K(s|\,x,y)$ \cite{CPTII,nnea} briefly reviewed above
-- there is no nonanalytic terms in $1/s$ like $\ln(1/s)$),
    \begin{equation}
     \Omega(s|\,x,y)=
     \Omega_0(x,y)+\frac1s\,\Omega_1(x,y)
     +O\left(\,\frac1{s^2}\,\right)           \label{3.2}
    \end{equation}
and obtain the series of recurrent equations for the coefficients
of this expansion
    \begin{eqnarray}
     &&F(\nabla)\,\Omega_0(x,y)=0,           \label{3.3}
     \\
     &&F(\nabla)\,\Omega_1(x,y)=
     (x-y)^\mu\nabla_\mu
     \Omega_0(x,y),                     \label{3.4}
     \\
     &&... \nonumber
    \end{eqnarray}

An obvious difficulty with the choice of the concrete solution for
this chain of equations is that they do not form a well posed
boundary value problem. Point is that natural zero boundary
conditions at spacetime infinity for the original kernel
$K(s|\,x,y)$ do not impose any boundary conditions on the function
$\Omega(s|\,x,y)$ except maybe the restriction on the growth of
$\Omega(s|\,x,y)$ to be slower than $\exp\,[+|x-y|^2/2s]$, because
of the exponential factor in (\ref{2.1}). On the other hand, this
freedom in choosing non-decreasing at $|x|\to\infty$ solutions
essentially helps to find them, because in the opposite case even
the existence of nontrivial solutions would be violated. Indeed
the elliptic equation (\ref{3.3}) with positive definite operator
$F(\nabla)$ (which we assume) would not have nonzero solutions
decaying at spacetime infinity. Thus, the only remaining criterion
for the selection of solutions in (\ref{3.3})-(\ref{3.4}) is the
requirement of symmetry of the coefficients $(\Omega_0(x,y),\,
\Omega_1(x,y),...)$ in their arguments. As we will see now, this
criterion taken together with certain assumptions of
\emph{naturalness} result in concrete solutions which will be
further checked on consistency by different methods including
perturbation theory, the variational equation for the heat kernel
trace and its metric analogue, etc.

The absence of falloff properties for the coefficients of the
expansion (\ref{3.2}) results in one more interesting peculiarity
of the late-time expansion. As we will now see, this expansion for
the functional {\it trace} of the heat kernel corresponding to
(\ref{3.2}) reads
    \begin{equation}
     {\Tr}K(s)=\frac1{(4\pi s)^{d/2}}\,
     \left\{\,s\,W_0+W_1
     +O\left(\,\frac1s\,\right)\,\right\},  \label{3.5}
    \end{equation}
which obviously implies that in spite of (\ref{1.11}) $W_n\neq\int
dx\:\Omega_n(x,x)$, $n=0,1,...$, because of the unit shift in the
power of $s$. The explanation of this visible mismatch between
(\ref{3.2}) and (\ref{3.5}) lies in the observation that the
$1/s$-expansion (\ref{3.2}) is not uniform in $x$ and $y$
arguments of $\,\Omega(s|\,x,y)$. Therefore, for fixed $s$ the
expression $\Omega(s|\,x,x)$ taken from this expansion fails to be
correct for $|x|\to\infty$ and, as a consequence, the heat kernel
functional trace (requiring integration up to spacetime infinity)
cannot be obtained by applying (\ref{1.11}) to (\ref{3.2}).

Fortunately, there is a means of circumventing this difficulty.
The functional trace of the heat kernel can be recovered from its
expansion due to the following variational equation
    \begin{equation}
     \frac{\delta \,{\Tr}K(s)}{\delta V(x)}
     =-sK(s|x,x),                            \label{3.6}
    \end{equation}
which is a direct corollary of the heat kernel definition. One
power of $s$ on the right hand side of this equation explains, in
particular, extra power of the proper time in (\ref{3.5}) as
compared to (\ref{3.2}). This equation reduces to the sequence of
variational equations
    \begin{eqnarray}
     \frac{\delta \,W_n}{\delta V(x)}
     =-\Omega_n(x,x), \qquad n=0,1,...\,,          \label{3.7}
    \end{eqnarray}
which will be used to obtain the first two coefficients in
(\ref{3.5}).

\subsection{Leading order}
\hspace{\parindent} The way the strategy outlined above works in
the leading order of the $1/s$-expansion is as follows
\cite{nnea}. Make a natural \emph{assumption} that $\Omega_0(x,y)$
at $|x|\to\infty$ is not growing and \emph{independent} of the
angular direction $n^\mu=x^\mu/|x|$ quantity $C(y)$ -- the
function of only $y$. Then the solution of Eq.(\ref{3.3}) subject
to boundary condition $\Omega_0(x,y)\,|_{\,|x|\to\infty}=C(y)$, is
unique and reads $\Omega_0(x,y)=\Phi(x)\,C(y)$, where $\Phi(x)$ is
a special function
    \begin{equation}
     \Phi (x)=1+\frac{1}{\Box-V}\,V(x)
     \equiv1+\int dy\:G(x,y)V(y)            \label{3.8}
    \end{equation}
solving the homogeneous equation subject to unit boundary
conditions at infinity
    \begin{eqnarray}
     \left\{\begin{array}{l}
     F(\nabla)\,\Phi(x)=0,\\
     \\
     \Phi (x)\rightarrow 1,
     \quad |x|\rightarrow\infty.
     \end{array}\right.                  \label{3.9}
    \end{eqnarray}
Then, the requirement of symmetry in $x$ and $y$ implies that
$\Omega_0(x,y)=C\,\Phi(x)\,\Phi(y)$, where the value of the
numerical normalization coefficient $C\!=\!1\,$ follows from the
comparison with the exactly known heat kernel in flat spacetime
with vanishing potential $V(x)=0$. Thus
    \begin{equation}
    \Omega_0(x,y)=\Phi(x)\,\Phi(y).     \label{3.10}
    \end{equation}

Note that with this result treated as valid up to infinity in $x$,
the heat kernel trace becomes badly defined because the integral
of the coincidence limit of $\Omega_0(x,y)$, is divergent
    \begin{eqnarray}
     \int dx\:\Omega(s|x,x)=\int
     dx\:\Phi^2(x)=\infty.             \nonumber
    \end{eqnarray}
On the contrary, the variational equation (\ref{3.7}) for $W_0$
satisfies the integrability condition in view of the relation
    \begin{eqnarray}
    \frac{\delta\Phi(x)}{\delta V(y)}
    =G(x,y)\,\Phi(y)                     \label{3.11}
    \end{eqnarray}
and has the following solution in the form of a well-defined
convergent integral \cite{nnea}
    \begin{equation}
    W_0=-\int dx\:V\,\Phi(x).           \label{3.12}
    \end{equation}

\subsection{Subleading order}
\hspace{\parindent} In the subleading order the equation
(\ref{3.4}) takes the form
    \begin{equation}
     F(\nabla)\,\Omega_1(x,y)=
     (x-y)^\mu\nabla_\mu
     \Phi(x)\Phi(y),                        \label{3.13}
    \end{equation}
with the inhomogeneous term on the right hand side, which is
slowly tending to zero at infinity in $|x|$ (and growing in
$|y|$). A symmetric in $x$ and $y$ solution that was found in
\cite{nneag}
    \begin{eqnarray}
     &&\Omega_1(x,y)=\frac1{\Box_x-V_x}\,(x-y)^\mu
     \nabla_\mu\Phi(x)\,\Phi(y)+(x\leftrightarrow y)\nonumber\\
     &&\qquad\qquad
     +\,2\,\frac1{\Box_x-V_x}\nabla_\mu\Phi(x)\,
     \frac1{\Box_y-V_y}
     \nabla^\mu\Phi(y),                      \label{3.14}
    \end{eqnarray}
has linearly growing in $x$ and $y$ terms\footnote{Here the label
of the differential operator in the denominator indicates on which
argument the corresponding Green's function is acting.}, which is
a corollary of the missing falloff property for the right hand
side of Eq.(\ref{3.13}). Thus, this solution is essentially
non-unique, but the correctness of its choice can be supported by
the fact that the corresponding coincidence limit $\Omega_1(x,x)$
guarantees the integrability condition of the variational equation
(\ref{3.7}) for $W_1$ \cite{nneag} and confirmed by the direct
summation of the perturbative series for ${\Tr}K(s)$ (see below
and Appendix A). The answer for $W_1$ reads
    \begin{eqnarray}
     W_1=\int dx\,\left\{1
     -2\,\nabla_\mu\Phi\,\frac1{\Box-V}\,
     \nabla^\mu\Phi\,\right\},        \label{3.15}
    \end{eqnarray}
where the nontrivial nonlocal term quadratic in gradients of
$\Phi(x)$ is given by a well-defined convergent integral in
distinction from the unit term which was added as a functional
integration constant following from the comparison of this result
with the exactly known case of a vanishing potential.

\subsection{Resummation of covariant perturbation theory}
\hspace{\parindent} To substantiate the nonperturbative algorithms
of this section and, in particular, to check the correct choice of
solutions for the coefficients of the $1/s$-expansion (\ref{3.2})
one can compare them with the result of the resummation of
nonlocal series in the perturbation technique of
\cite{CPTI,CPTII,CPTIII}. In this technique the functional trace
of the heat kernel is expanded as nonlocal series
(\ref{CPT1})-(\ref{Kn}) in powers of the potential $V$ with
explicitly calculable coefficients - nonlocal form factors
$F_{n}(s|\,x_{1},x_{2},...x_{n})$. Their leading asymptotic
behavior at large $s$ was obtained in \cite{CPTII}, and it can
also be extended to the first subleading order in $1/s$
\cite{nnea}. Then in this approximation one can explicitly perform
infinite summation of power series in the potential to confirm the
answers for $W_0$ and $W_1$ obtained above.

According to \cite{CPTII} the heat kernel trace is local in the
first two orders of the perturbation theory (\ref{CPT1})
    \begin{eqnarray}
    &&{\Tr}K_{0}(s)
    =\frac{1}{(4\pi s)^{d/2}}\int dx,  \label{3.16}\\
    &&{\Tr}K_{1}(s)
    =-\frac{s}{(4\pi s)^{d/2}}\int dx\,V(x),   \label{3.17}
    \end{eqnarray}
and in higher orders it reads as
    \begin{equation}
    {\Tr}K_{n}(s)=\frac{(-s)^{n}}{(4\pi s)^{d/2}n}
    \int dx\,\big<e^{s\Omega_{n}}\big>\,
    V(x_{1})V(x_{2})...V(x_{n})
    \Big|_{x_{1}=...=x_{n}=x},\quad n\geq 2.  \label{B1}
    \end{equation}
Here $\Omega _{n}$ is a differential operator acting on the
product of $n$ potentials
    \begin{equation}
    \Omega_{n}=\sum\limits_{i=1}^{n-1}
    \nabla_{i+1}^{2}+2\sum\limits_{i=2}^{n-1}
    \sum\limits_{k=1}^{i-1}\beta_{i}(1-\beta_{k})
    \nabla_{i+1}\nabla_{k+1},                    \label{B2}
    \end{equation}
expressed in terms of the partial derivatives labelled by the
indices $i$ implying that $\nabla _{i}$ acts on $V(x_{i})$. It is
assumed in (\ref{B1}) that after the action of all derivatives on
the respective terms all $x_{i}$ are set equal to $x$. It is also
assumed that the spacetime indices of all derivatives $\nabla
=\nabla ^{\mu }$ are contracted in their bilinear combinations,
$\nabla _{i}\nabla _{k}\equiv \nabla _{i}^{\mu }\nabla _{\mu\,k}$.
The differential operator (\ref{B2}) depends on the parameters
$\beta _{i}$, $\,i=1,...,n\!-\!1$, which are defined in terms of
the parameters $\alpha_{i}$, $\,i=1,...,n$ as
    \begin{equation}
    \beta _{i}=\alpha_{i+1}+\alpha_{i+2}+...+\alpha_{n},
    \end{equation}
and the angular brackets in $\big<e^{s\Omega _{n}}\big>$ imply
that this operator exponent is integrated over compact domain in
the space of $\alpha$-parameters
    \begin{equation}
    \big<e^{s\Omega_{n}}\big>\equiv
    \int\limits_{\alpha_{i}\geq 0}d^{n}\alpha
    \,\delta \Big(\sum_{i=1}^{n}\alpha_{i}-1\Big)\,
    \exp (s\Omega_{n}).                       \label{3.18}
    \end{equation}

The late time behavior of $\,{\Tr} K_{n}(s)\,$ is thus determined
by the asymptotic behavior of this integral at
$s\rightarrow\infty$, which can be calculated using the Laplace
method. To apply this method, let us note that $\Omega_{n}$ is a
\emph{negative} semidefinite operator (this is shown in the
Appendix B of \cite{CPTII}) which degenerates to zero at $n$
points of the integration domain: $(0,...,0,\alpha
_{i}\!=\!1,0,...,0)$, $\,i=1,...,n$. Therefore the asymptotic
expansion of this integral is given by the contribution of the
corresponding $n$ maxima of the integrand at these points. The
integration by parts in (\ref{B1}) is justified by the formal
identity $\nabla_{1}+\nabla_{2}+...+\nabla_{n}=0$ . Using it one
can show that the contributions of all these maxima are equal, so
that it is sufficient to calculate only the contribution of the
point $\alpha_{1}=1$, $\,\alpha_{i}=0$, $\,i=2,...,n$. In the
vicinity of this point it is convenient to rewrite the expression
for $\Omega_{n}$ in terms of the independent $(n-1)$ variables
$\alpha_{2},\alpha_{3},...,\alpha_{n}$, the remaining
$\alpha_{1}=1-\sum_{i=2}^{n}\alpha_{i}$,
    \begin{equation}
    \Omega_{n}=\sum_{i=2}^{n}\alpha_{i}D_{i}^{2}
    -\sum_{m,k=2}^{n}\alpha_{m}\alpha_{k}D_{m}D_{k},\label{3.19}
    \end{equation}
where the operator $D_{m}$ is defined as
    \begin{equation}
    D_{m}=\nabla_{2}+\nabla_{3}+...+\nabla_{m},
    \quad m=2,...,n.                              \label{D}
    \end{equation}

The details of the further calculation are presented in the
Appendix A. Here we only specify the structure of the result in
the $n$-th order of perturbation theory, which shows that one can
perform explicit summation of nonlocal perturbation series leading
to (\ref{3.12}) and (\ref{3.15}). This structure is most
characteristic in the leading order approximation. The $n$-th
order term reads
    \begin{eqnarray}
    &&{\Tr}\,K_{n}(s)=\frac{1}{(4\pi s)^{d/2}}\int dx\,
    \left[ -s\frac{1}{D_{2}^{2}...D_{n}^{2}}
    +O\left( s^0\right) \right] \,V_{1}V_{2}...V_{n}
    \nonumber\\
    &&\qquad\qquad=-\frac{s}{(4\pi s)^{d/2}}
    \int dx\,\underbrace{V\frac{1}{\Box }V\frac{1}{\Box }...
    V\frac{1}{\Box }}_{n-1}\,V(x)
    +O\left( \frac{1}{s^{d/2}}\right),   \label{3.20}
    \end{eqnarray}
where in view of the definition of the generalized derivative
(\ref{D}) the action of the nonlocal operator
$1/D_{2}^{2}...D_{n}^{2}$ on the multi-point product
$V_{1}V_{2}...V_{n}$ is rewritten as the $(n-1)$-th power of the
nonlocal operator $V\times1/\Box$ \emph{acting to the right} on
the first power of $V$. Thus the expression (\ref{3.20}) turns out
to be the term of geometric progression in powers of $V(1/\Box)$,
and the summation yields
    \begin{eqnarray}
    &&{\Tr}K(s)=\frac{1}{(4\pi s)^{d/2}}
    \int dx\,\Big\{-s\,\Box \frac{1}{\Box-V}\,V(x)
    +O(s^{0})\Big\}
    \nonumber\\
    &&\qquad\qquad=\frac{1}{(4\pi s)^{d/2}}
    \int dx\,\Big\{-s\,V\Phi (x)+O(s^{0})\Big\}.    \label{3.21}
    \end{eqnarray}
Similarly, in the subleading order the calculation reduces to the
summation of multiple -- duplicate and triplicate -- geometric
progressions in powers of the same nonlocal operator. This
summation gives
    \begin{equation}
    {\Tr}K(s)=\frac1{(4\pi s)^{d/2}}
    \int dx\,\Big\{-sV\Phi+1
    -2\nabla_\mu\Phi\frac1{\Box-V}\nabla^\mu\Phi
    +O\left(s^{-1}\right)\Big\}        \label{3.22}
    \end{equation}
and, thus, fully confirms Eqs.(\ref{3.12}) and (\ref{3.15})
obtained by another essentially nonperturbative method.

\section{New nonlocal effective action  }
\hspace{\parindent} Nonperturbative asymptotics of the heat kernel
allows one to improve essentially the calculation of the effective
action. The new approximation which unifies the knowledge of both
the early-time and late-time behaviors of ${\Tr}K(s)$ incorporates
both the ultraviolet and infrared properties of the theory and
generates new structures in the effective action. These structures
generalize the Coleman-Weinberg local potential (\ref{2.11}) to
physical situations when the field is not homogeneous throughout
the whole spacetime, but rather tends to zero at infinity. As we
saw in Sect.2.2 only for extremely slow and physically
uninteresting falloff (\ref{potfall}) with $p<2$ the deviation
from homogeneity can be treated by perturbations. For a faster
decrease at $|x|\to\infty$ the infrared cutoff of the proper-time
integral fails within the modified gradient expansion, and we have
to use either the nonlocal perturbation theory of Sect.2.3 or the
nonperturbative asymptotics of Sect.3. The latter allows one to
obtain both nonlocal and nonperturbative action which captures in
a nontrivial way the edge effects of a transition domain between
the spacetime interior at finite $|x|$ to vanishing potential at
$|x|\to\infty$.

The key idea to build this new approximation is to replace
${\Tr}K(s)$ in (\ref{1.10}) by some approximate function
${\Tr}\bar{K}(s)$ such that both early and late time asymptotics
are satisfied and the integral over $s$ is explicitly calculable.
Here we exploit the simplest possibility -- namely, take two
simple functions ${\Tr}{K}_{<}(s)$ and ${\Tr}{K}_{>}(s)$
    \begin{eqnarray}
     &&{\Tr}K_{<}(s)=\frac{1}{(4\pi s)^{2}}
     \int dx\,e^{-sV},                     \label{4.1}\\
     &&{\Tr}K_{>}(s)=\frac{1}{(4\pi s)^{2}}
     \int dx\,(1-sV\Phi),                    \label{4.2}
    \end{eqnarray}
which coincide with the leading asymptotics of ${\Tr}K(s)$ at
$s\rightarrow 0$ (modified gradient expansion with derivative
terms omitted) and $s\rightarrow \infty$ and use them to
approximate ${\Tr}K(s)$ respectively at $0\leq s\leq s_{\ast}$ and
$s_{\ast }\leq s<\infty$ for some $s_{\ast }$
    \begin{eqnarray}
     \Tr \bar{K}(s)=\left\{\begin{array}{l}
     \Tr K_{<}(s),\quad s<s_{\ast},\\
     \\
     \Tr K_{>}(s),\quad s>s_{\ast}.
     \end{array}\right.              \label{4.3}
    \end{eqnarray}
The value of $s_{\ast}$ will be determined from the requirement
that these two functions match at $s_{\ast}$, which will guarantee
the stationarity of $\bar\SGamma$ with respect to the choice of
$s_{\ast}$, $\partial \bar{\SGamma}/\partial s_{\ast}=0$,
    \begin{equation}
     \Tr K_{<}(s_{\ast})=\Tr K_{>}(s_{\ast}).  \label{4.4}
    \end{equation}

Thus, the new approximation for the action reads as
    \begin{equation}
    \bar\SGamma=
    -\frac{1}{2}\int\limits_{0}^{s_{\ast} }
    \frac{ds}{s}{\Tr}K_{<}(s)
    -\frac{1}{2}\int\limits_{s_{\ast }}^{\infty }
    \frac{ds}{s} {\Tr}K_{>}(s),                   \label{4.5}
    \end{equation}
and its deviation from the exact $\SGamma_\mathrm{1-loop}$
proportional to ${\Tr}[K(s)-\bar{K}(s)]$ can then be treated by
perturbations. This piecewise-smooth approximation is efficient
only if the ranges of validity of two asymptotic expansions
(respectively for small and big $s$) overlap with each other and
the point $s_{\ast}$ belongs to this overlap. In this case the
corrections due to the deviation of ${\Tr}\bar{K}(s)$ from the
exact ${\Tr}K(s)$ are uniformly bounded everywhere and one can
expect that (\ref{4.5}) would give a good zeroth-order
approximation to an exact result. This requirement can be
satisfied at least for two rather wide classes of potentials
$V(x)$. They have finite amplitude $V_0$ within their compact
support of size $R$ \cite{nnea},
    \begin{eqnarray}
     &&V(x)=0,\quad |x|\geq R,                 \nonumber\\
     &&V(x)\sim V_0,\quad |x|\leq R,           \label{4.6}
    \end{eqnarray}
and have the property that their derivatives are not too high and
uniformly bounded by the quantity of the order of magnitude
$V_0/R$.

One class is when the potential is small in units of the inverse
size of its compact support
    \begin{equation}
    V_0 R^2\ll 1.                 \label{4.7}
    \end{equation}
Consider first the case of four-dimensional spacetime, $d=4$.
Simple calculation presented in Appendix B yields in this case the
following answer \cite{nnea} for the finite part of the action,
which is valid up to corrections proportional to this smallness
parameter
    \begin{eqnarray}
    &&\bar{\SGamma}\simeq \frac{1}{64\pi ^{2}}
    \int d^4x\,V^{2}(x)\,\ln \Big(\int d^4y\,V^{2}(y)\Big)\nonumber\\
    &&\qquad\qquad-
    \frac{1}{64\pi ^{2}}\int d^4x\,V^{2}(x)
    \ln \Big(\int d^4y\,V\frac{\mu ^{2}}
    {V-\Box }V(y)\Big)\;,\quad d=4.                \label{4.8}
    \end{eqnarray}
In what follows we disregard the ultraviolet divergent part of the
action and absorb all finite renormalization type terms $\sim\int
dx\,V^{d/2}$ in the redefinitions of $\mu^2$.

Note, that this renormalization mass parameter $\mu^{2}$ makes the
argument of the second logarithm dimensionless and plays the same
role as for the Coleman-Weinberg potential. However, the original
Coleman-Weinberg term for small potentials of the type (\ref{4.7})
gets replaced by the other qualitatively new nonlocal structure.
For small potentials spacetime gradients dominate over their
magnitude and, therefore, the Coleman-Weinberg term does not
survive in this approximation. Still it can be recovered in the
formal limit of the constant potential, when the argument of the
second logarithms tends to $\mu^2\int dx\,V$ and the infinite
volume factor ($\int dx$) gets cancelled in the difference of two
logarithms,
    \begin{equation}
    \bar{\SGamma}\to\SGamma_\mathrm{CW}\equiv
    \frac{1}{64\pi^{2}}
    \int d^4x\,V^{2}\ln\frac{V}{\mu ^{2}},\quad
    V(x)\to\mathrm{const}.                         \label{4.9}
    \end{equation}
This justifies the consistency of this approximation.

As shown in Appendix B, for higher (even) dimensions $d>4$ the
logarithmic term of the action turns out to be subleading, and the
answer is dominated by the following renormalization-independent
and \emph{negative-definite} part of the expression
    \begin{eqnarray}
    &&\bar{\SGamma}\simeq
    -\frac{1}{(8\pi)^{d/2}}\,\frac2{d\,(d-2)}
    \,\int dx\,V^{2}(x)\,\left(\frac{\int dy\;V^2(y)}
    {\vphantom{\displaystyle\frac{a}a}\int dy\,V\!
    \frac1{V-\Box}V(y)}\right)^{d/2-2}   \nonumber\\
    &&\qquad\qquad\qquad
    +\frac1{2\,(4\pi)^{d/2}}\int dx\,\frac{(-V)^{d/2}(x)}{(d/2)!}\,
    \ln\!\left(\!\frac{\int dy\;V^2(y)}
    {\vphantom{\displaystyle\frac{a}a}\int dy\,V\!
    \frac{\,\displaystyle\mu^2\!}
    {V-\Box}V(y)}\right)\;.                     \label{4.8a}
    \end{eqnarray}
The second term here for small but spatially variable potential is
generally much smaller than the first one, because within the
bound (\ref{4.7}) (see Appendix B)
    \begin{eqnarray}
    \frac{\int dy\;V^2(y)}
    {\vphantom{\displaystyle\frac{a}a}\int dy\,V\!
    \frac1{V-\Box}V(y)}\gg V(x)\;.                 \label{4.8b}
    \end{eqnarray}
However, in the limit of constant potential this ratio (which is
actually $1/s_\ast$) tends to $V$, so that both terms become of
the same order of magnitude. The first term then turns out to be
of a purely renormalization nature $\sim\int dx\,V^{d/2}$, while
the second term goes over into the (logarithmically stronger)
$d$-dimensional Coleman-Weinberg potential (\ref{2.11}). So the
correspondence principle holds also in higher dimensions.

Another class of potentials, when the piecewise smooth
approximation is effective, corresponds to the opposite limit,
    \begin{equation}
    V_0 R^2\gg 1,              \label{4.10}
    \end{equation}
that is big potentials in units of the inverse size of their
support. In this case spacetime gradients do not dominate the
amplitude of the potential and the calculation in Appendix B shows
that the effective action contains the Coleman-Weinberg term
modified by the special nonlocal correction (previously derived
for $d=4$ in \cite{nnea})
    \begin{equation}
    \bar{\SGamma}\simeq\SGamma_\mathrm{CW}
    +\frac{1}{(4\pi)^{d/2}}\,
    \frac{2}{d\,(d-2)}\,\int\limits_{|x|\leq R}
    \!dx\,\big<V\Phi \big>^{d/2}.          \label{4.11}
    \end{equation}
This correction involves the average value of the function
$V\Phi(x)$ on the compact support domain
    \begin{equation}
    \big<V\Phi \big>\equiv
    \frac{\int\limits_{|x|\leq R}\!dx\,V\Phi }
    {\int\limits_{|x|\leq R}\!dx}.           \label{4.12}
    \end{equation}
Again this algorithm correctly stands the formal limit of a
constant potential, because in this limit the function $\Phi(x)$
given by (\ref{3.8}) formally tends to zero (and the size $R$
grows to infinity).

\section{Inclusion of gravity}
\hspace{\parindent} Remarkable property of the obtained late-time
asymptotics is that it can be nearly straightforwardly generalized
to curved spacetime. In this case the flat metric gets replaced by
the curved one and the interval in the ansatz (\ref{2.1}) goes
over to the world function -- one half of the geodesic distance
squared between the points $x$ and $y$
    \begin{eqnarray}
     &&\delta_{\mu\nu}
      \rightarrow g_{\mu\nu}(x), \nonumber\\
     &&\frac{|x-y|^2}{2}\rightarrow\sigma(x,y). \label{5.1}
    \end{eqnarray}
The semiclassical ansatz (\ref{2.1}) takes the form
    \begin{eqnarray}
     &&K(s|\,x,y)=\frac1{(4\pi s)^{d/2}}\,
     \exp\left[-\frac{\sigma(x,y)}{2s}\right]\,
     \Omega(s|\,x,y)\,g^{1/2}(y),                \label{5.2}
    \end{eqnarray}
where $\Omega(s|\,x,y)$ is a biscalar quantity which instead of
(\ref{2.2}) has a small-time limit
    \begin{eqnarray}
     &&\Omega(s|\,x,y)\;
     \rightarrow\;\Delta^{1/2}(x,y),\quad s\to 0,
     \nonumber\\
     &&\vphantom{\frac11}\Delta(x,y)=g^{-1/2}(x)
     \left(\det{\partial_\mu^x\partial_\nu^y
       \sigma(x,y)}\right)
     g^{-1/2}(y)\neq 0                         \label{5.3}
    \end{eqnarray}
in terms of the (dedensitized) Pauli-Van Vleck-Morette determinant
\cite{PhysRep,DeWitt}\footnote{We define the
$\delta(x,y)$-function as a scalar with respect to the first
argument $x$ and as a density of unit weight with respect to the
second one -- $y$. Correspondingly the heat kernel has the same
weights of their arguments. This asymmetry in $x$ and $y$ explains
the presence of the factor $g^{1/2}(y)$ in (\ref{5.2}) and a
biscalar nature of $\Omega(s|\,x,y)$.}.

Disentangling of $\Delta^{1/2}(x,y)$ as a separate factor in
(\ref{5.1}) is not useful for the purposes of late time expansion.
However, the quantity is rather important and related to a serious
simplifying assumption which underlies our results. The assumption
we make is the absence of focal points in the congruence of
geodesics determining the world function $\sigma(x,y)$. We assume
that for all pairs of points $x$ and $y$, $\Delta(x,y)\neq 0$,
which guarantees that $\sigma(x,y)$ is globally and uniquely
defined on the asymptotically-flat spacetime in question. This
assumption justifies the ansatz (\ref{5.2}) which should be
globally valid because the coefficients of the expansion
(\ref{3.2}) will satisfy elliptic boundary-value problems with
boundary conditions at infinity.

We also assume that the metric is asymptotically flat and has at
spacetime infinity in cartesian coordinates the following falloff
behavior characteristic of $d$-dimensional Euclidean spacetime
    \begin{eqnarray}
    g_{\mu\nu}(x)\,\Big|_{\,|x|\to\infty}
    =\delta_{\mu\nu}
    +O\left(\frac1{|x|^{d-2}}\right).    \label{5.4}
    \end{eqnarray}
This requirement does not exclude caustics in the geodesic flow,
$\Delta(x,y)=0$, which depend on local properties of the
gravitational field, unrelated to its long-distance behavior. The
assumption of geodesic convexity might be too strong to
incorporate physically interesting situations, but we believe that
the late-time asymptotics will survive in the presence of caustics
(though, maybe by the price of additional contributions which go
beyond the scope of this paper)\footnote{This hope is based on a
simple fact that the leading order of the $1/s$-expansion is not
sensitive to the properties of the world function at all. Beyond
this order the main object of interest, ${\Tr}K(s)$, involves the
coincidence limit of the world function $\sigma(x,x)=0$, while its
asymptotic coefficients in $\Omega_n(x,y)$ nonlocally depend on
global geometry and can acquire from caustics additional
contributions analogous to those of multiple geodesics connecting
the points $x$ and $y$ beyond the geodesically convex neighborhood
\cite{camporesi}.}.

\subsection{Leading order}
\hspace{\parindent} Repeating the arguments of Sect.3 one
immediately finds the sequence of recurrent equations for the
coefficients of the $1/s$-expansion (\ref{3.2}) for the operator
$F(\nabla)$, (\ref{1.5})-(\ref{1.6}), in curved spacetime. Their
solution in the leading order turns out to be a straightforward
covariantization of the flat-space result (\ref{3.10}) with the
universal scalar function $\Phi(x)$ being now a functional of both
metric and potential,
    \begin{equation}
     \Phi(x)\equiv\Phi(x)[g_{\mu\nu},V]=
     1+\int dy\: G(x,y)V(y),               \label{5.5}
    \end{equation}
in terms of the curved-space Green's function $G(x,y)\equiv
G(x,y)[V,g_{\mu\nu}]$ subject to the same Dirichlet boundary
conditions at infinity
    \begin{eqnarray}
     \left\{ \begin{array}{l}
     F(\nabla)\,G(x,y)=\delta(x,y), \\
     G(x,y)|_{_{|x|\to\infty}}=0
     \end{array} \right.                  \label{5.6}
    \end{eqnarray}

Thus, the leading order of late-time expansion for $K(s|\,x,y)$ is
just a direct covariantization of the flat-space result.
Remarkably, almost the same situation holds for the functional
trace. Its leading order is given by two terms. One of them is a
straightforward covariantization of (\ref{3.13}) and another
represents a new structure -- the surface integral over spacetime
infinity reflecting the asymptotically-flat properties of its
metric.
    \begin{eqnarray}
     &&W_0=-\int dx\,g^{1/2}\,V\,\Phi(x)
     +\frac16\,\Sigma[\,g_\infty\,],    \label{5.7}\\
     &&\Sigma[\,g_\infty\,]
     =\int\limits_{|x|\to\infty}\!
     d\sigma^\mu\;\delta^{\alpha\beta}
     \Big(\partial_\alpha
     g_{\beta\mu}-\partial_\mu g_{\alpha\beta}\Big). \label{5.8}
    \end{eqnarray}

The proof of this statement cannot rely on the covariant expansion
in powers of $V$ and $R_{\mu\nu}$
\cite{CPTI,CPTII,CPTIII,asymp,basis}. This is because, in contrast
to the perturbation theory in potential, the $n$-th term of this
curvature expansion like in (\ref{B1}) is not explicitly available
(even the calculation of the third order presents enormous
calculational problem \cite{CPTIII}). Therefore, a metric analogue
of the variational equation (\ref{3.7}) should be used to recover
the functional trace from the kernel $K(s|\,x,y)$. But even this
does not turn out to be sufficient, because the variational
equation (\ref{3.7}) is valid, strictly speaking, only up to
surface terms, which do not vanish in the case of the metric
variation. Indeed, the variational equation for the operator
exponent actually reads
    \begin{equation}
     \delta_g \Tr e^{sF}=\int_0^s dt\:
     \Tr\left[e^{(s-t)F}\delta_g F e^{tF}\right]
     =s \Tr\left[\delta_gF e^{sF}\right]
     +\mathrm{surface\;\; terms},               \label{5.9}
    \end{equation}
where the second equality holds due to the cyclic property of the
functional trace. Unlike for finite-dimensional matrices, here the
cyclic property of the infinite-dimensional functional trace is
based on multiple (actually infinitely multiple) integration by
parts and, therefore, can bring nonvanishing surface integrals.
Fortunately, for asymptotically-flat spacetime they can be
completely taken into account within the first order of
perturbation theory. Thus, for the derivation of (\ref{5.7}) we
will use the combination of the variational and perturbation
techniques.

Consider first the metric variational derivative of ${\Tr}K(s)$ in
the class of variations $\delta g_{\mu\nu}$ sufficiently rapidly
decaying at spacetime infinity, so that no surface terms are
arising while integrating by parts in the expression above. The
corresponding variational derivative then takes the form
    \begin{eqnarray}
     &&\frac{\delta{\Tr} K(s)}
     {\delta g_{\mu\nu}(x)}=
     s\int dy\,\frac{\delta F(\nabla_y)}{\delta g_{\mu\nu}(x)}
     \,K(s|y,y')\Big|_{y'=y}
     \nonumber\\
     &&\qquad\qquad\qquad\qquad\qquad=-s\,g^{1/2}(x)\,
     f^{\mu\nu}(\nabla_x,\nabla_y)
     K(s|x,y)\,\Big|_{x=y},       \label{5.10}
    \end{eqnarray}
where
    \begin{eqnarray}
    f^{\mu\nu}(\nabla_x,\nabla_y)=
    -\nabla_x^{(\mu}\nabla_y^{\nu)}
    +\frac12\,g^{\mu\nu}\Box_x
    +\frac12\,g^{\mu\nu}
    \nabla^\lambda_x\nabla_\lambda^y.            \label{5.11}
    \end{eqnarray}
Direct check then shows that in this class of variations which
probe only the bulk part of (\ref{5.7}) Eq.(\ref{5.10}) is
satisfied in the leading order of the $1/s$-expansion \cite{nneag}
    \begin{eqnarray}
     \frac{\delta W_0}{\delta g_{\mu\nu}(x)}=
     -g^{1/2}(x)\,f^{\mu\nu}(\nabla_x,\nabla_y)\,
     \Phi(x)\Phi(y)\,\Big|_{x=y}\,.             \label{5.11a}
    \end{eqnarray}
The variation of $W_0$ given by Eq.(\ref{5.7}) is based here on
the metric variational derivative of $\Phi(x)$ -- the analogue of
Eq.(\ref{3.11})
  \begin{eqnarray}
  \frac{\delta\Phi(x)}{\delta g_{\mu\nu}(y)}=G(x,y)\,
  f^{\mu\nu}(\stackrel{\rightarrow}{\nabla}_y,
  \stackrel{\leftarrow}{\nabla}_y)
  \,\Phi(y),                                    \label{5.12}
  \end{eqnarray}
where the arrow indicates in which direction the corresponding
derivative is acting.

Metric variations decaying at $|x|\to\infty$ as
asymptotically-flat corrections (\ref{5.4}), $\delta
g_{\mu\nu}(x)\sim 1/|x|^{d-2}$, induce nonvanishing surface terms
in the variational equation (\ref{5.9}) and, therefore, they
cannot be checked with the use of (\ref{5.10}). Fortunately, the
covariant perturbation theory of Sect.2.3 shows \cite{CPTII} that
these terms (arising from cyclic permutations of heat kernels
under the sign of the functional trace) appear only in the linear
order of perturbation expansion in $h_{\mu\nu}$ -- deviation of
the metric from the flat-space one. Therefore, they can be
recovered by simply comparing (\ref{5.7}) with the late time
asymptotics obtained in \cite{CPTIII,asymp} to cubic order in
curvature and potential\footnote{This approximation is more than
sufficient to check the term linear in metric perturbation
$h_{\mu\nu}$. In covariant perturbation theory this perturbation
gets expanded in the infinite power series in curvatures -- this
is the price one pays for having this expansion generally
covariant.}. This asymptotics for the operator of the form
(\ref{1.5})-(\ref{1.6}) reads
    \begin{eqnarray}
    \Tr K(s)\,&=&\,-\frac{s}{(4\pi s)^{d/2}}\,\int
    dx\,g^{1/2}\,\left\{\,V+V\frac1\Box\,V+V\frac1\Box\,
    V\frac1\Box\,V+O\left(V^4\right)\right\}\nonumber\\
    &&+\;\frac16\frac{s}{(4\pi s)^{d/2}}\,\int dx\,g^{1/2}\,\left\{\,R
    -\,R_{\mu\nu}\frac1{\Box} R^{\mu\nu}
    +\frac12\,R\frac1{\Box} R\right.\nonumber\\
    &&\qquad\qquad\qquad
       +\frac12\,R\left(\frac1{\Box}
    R^{\mu\nu}\right)\frac1{\Box} R_{\mu\nu}
       -R^{\mu\nu}\left(\frac1{\Box}
    R_{\mu\nu}\right)\frac1{\Box} R
       \nonumber\\
       &&\qquad\qquad\qquad
       +\left(\frac1{\Box} R^{\alpha\beta}\right)
    \left(\nabla_\alpha\frac1{\Box} R\right)
    \nabla_\beta\frac1{\Box} R\nonumber\\
    &&\qquad\qquad\qquad
       -2\,\left(\nabla^\mu\frac1{\Box} R^{\nu\alpha}\right)
    \left(\nabla_\nu\frac1{\Box}
    R_{\mu\alpha}\right)\frac1{\Box} R \nonumber\\
    &&\qquad\qquad\qquad
    -\left.2\,\left(\frac1{\Box} R^{\mu\nu}\right)
    \left(\nabla_\mu\frac1{\Box}
    R^{\alpha\beta}\right)\nabla_\nu\frac1{\Box}
    R_{\alpha\beta}
    +\mathrm{O}\,[\,R_{\mu\nu}^4\,]\,\right\}.    \label{5.13}
    \end{eqnarray}
The nonlocal expansion of $\Phi(x)$
    \begin{eqnarray}
    \Phi(x)=1+\frac1\Box\,V(x)+\frac1\Box\,
    V\frac1\Box\,V(x)+O\left(V^3\right)        \label{5.14}
    \end{eqnarray}
obviously recovers from the first integral here the first term of
(\ref{5.7}) explicitly containing only powers of potential with
\emph{metric-dependent} nonlocalities.

The second integral in (\ref{5.13}) is a topological invariant
independent of local metric variations in the interior of
spacetime -- exactly in this class of $\delta g_{\mu\nu}(x)$ the
functional derivative of (\ref{5.10}) was calculated above. Direct
expansion in powers of $h_{\mu\nu}$,
$\,g_{\mu\nu}=\delta_{\mu\nu}+h_{\mu\nu}$, on flat-space
background in cartesian coordinates shows that this term reduces
to the surface integral at spacetime infinity. For the class of
asymptotically flat metrics with $h_{\mu\nu}(x)\sim 1/|x|^{d-2}$,
$\,|x|\to\infty$, this surface integral is linear in perturbations
(contributions of higher powers of $h_{\mu\nu}$ to this integral
vanish) and involves only a \emph{local} asymptotic behavior of
the metric $g^\infty_{\mu\nu}(x)=\delta_{\mu\nu}
+h_{\mu\nu}(x)\,\Big|_{\,|x|\to\infty}$,
    \begin{eqnarray}
    \int dx\,g^{1/2}\,&&\left\{\,R
    -\,R_{\mu\nu}\frac1{\Box} R^{\mu\nu}
    +\frac12\,R\frac1{\Box} R\right.\nonumber\\
       &&+\frac12\,R\left(\frac1{\Box}
    R^{\mu\nu}\right)\frac1{\Box} R_{\mu\nu}
       -R^{\mu\nu}\left(\frac1{\Box}
    R_{\mu\nu}\right)\frac1{\Box} R\nonumber\\
       &&
       +\left(\frac1{\Box} R^{\alpha\beta}\right)
    \left(\nabla_\alpha\frac1{\Box} R\right)
    \nabla_\beta\frac1{\Box} R\nonumber\\
    &&-2\,\left(\nabla^\mu\frac1{\Box} R^{\nu\alpha}\right)
    \left(\nabla_\nu\frac1{\Box}
    R_{\mu\alpha}\right)\frac1{\Box} R \nonumber\\
    &&
    -\left.2\,\left(\frac1{\Box} R^{\mu\nu}\right)
    \left(\nabla_\mu\frac1{\Box}
    R^{\alpha\beta}\right)\nabla_\nu\frac1{\Box}
    R_{\alpha\beta}
    +\mathrm{O}[\,R_{\mu\nu}^4\,]\,\right\}\nonumber\\
    &&\nonumber\\
    &&\qquad\qquad\qquad\qquad\qquad
    =\int\limits_{|x|\to\infty} d\sigma^\mu\,
    \big(\partial^\nu
    h_{\mu\nu}-\partial_\mu h\Big)
    \equiv\Sigma\,[\,g_\infty\,].        \label{5.15}
    \end{eqnarray}
Here $d\sigma^\mu$ is the surface element on the sphere of radius
$|x|\to\infty$, $\,\partial^\mu=\delta^{\mu\nu}\partial_\nu$ and
$h=\delta^{\mu\nu}h_{\mu\nu}$. Covariant way to check this
relation is to calculate the metric variation of this integral and
show that its integrand is the total divergence which yields the
surface term of the above type linear in $\delta
g_{\mu\nu}(x)=h_{\mu\nu}(x)$. This is explicitly done in Appendix
B. Thus, the correct expression for $W_0$ is indeed modified by
the the surface integral $\Sigma\,[\,g_\infty\,]$, and this
integral does not contribute to the metric variational derivative
$\delta W_0/\delta g_{\mu\nu}(x)$ at finite $|x|$.

For asymptotically-flat metrics with a power-law falloff at
infinity $h_{\mu\nu}(x)\sim M/|x|^{d-2}$, $\,|x|\to\infty$, the
surface integral $\Sigma\,[\,g_\infty]$ forms the contribution to
the Einstein action
    \begin{eqnarray}
    S_\mathrm{E}[\,g\,]\equiv -\int dx\,g^{1/2}\,R(g)
      +\Sigma\,[\,g_\infty],     \label{5.17}
    \end{eqnarray}
which guarantees the correctness of the variational procedure
leading to Einstein equations. Covariantly this integral can also
be rewritten in the Gibbons-Hawking form
$S_{GH}\,[\,g\,]=\Sigma\,[\,g_\infty]$ -- the double of the
extrinsic curvature trace $K$ on the boundary (with a properly
subtracted infinite contribution of the flat-space background)
\cite{GH}
    \begin{eqnarray}
    \Sigma\,[\,g_\infty]=-2\int_\infty\!
    d^{d-1}\sigma\,\Big(g^{(d-1)}\Big)^{1/2}\,
    \Big(K-K_0\Big).                    \label{5.16}
    \end{eqnarray}
Thus, this is the surface integral of the \emph{local} function of
the boundary metric and its normal derivative. The virtue of the
relation (\ref{5.15}) is that it expresses this surface integral
in the form of the spacetime (bulk) integral of the
\emph{nonlocal} functional of the bulk metric. The latter does not
explicitly contain auxiliary structures like the vector field
normal to the boundary, though these structures are implicitly
encoded in boundary conditions for nonlocal operations in the bulk
integrand of (\ref{5.15}).

Note also, in passing, that the relation (\ref{5.15}) can be used
to rewrite the (Euclidean) Einstein-Hilbert action (\ref{5.17}) as
the \emph{nonlocal} curvature expansion which begins with the
\emph{quadratic} order in curvature. This observation serves as a
basis for covariantly consistent nonlocal modifications of
Einstein theory \cite{nonlocal} motivated by the cosmological
constant and cosmological acceleration problems \cite{A-HDDG}.

\subsection{Conformal properties}
\hspace{\parindent} Important case of the \emph{metric-dependent}
potential in (\ref{1.5})-(\ref{1.6}) corresponds to the operator
of the conformal scalar field. In $d$ dimensions it reads
    \begin{eqnarray}
      F(\nabla)=\Box-\frac14\,\frac{d-2}{d-1}\,R.  \label{5.18}
    \end{eqnarray}
Under local conformal (Weyl) transformations of the metric tending
to identity at infinity
    \begin{eqnarray}
      &&\bar{g}_{\mu\nu}(x)=
      \Psi^{4/(d-2)}(x)\,g_{\mu\nu}(x),\nonumber\\
      &&\Psi(x)=
      1+O\left(\frac1{|x|^{d-2}}\right),
      \quad |x|\to\infty            \label{5.19}
    \end{eqnarray}
this operator transforms homogeneously by multiplication from the
left and from the right with certain powers of the conformal
factor
    \begin{eqnarray}
      \bar{F}(\nabla)=
      \Psi^{-1-4/(d-2)}(x)\,F(\nabla)\,\Psi(x).   \label{5.20}
    \end{eqnarray}

In virtue of this property, the universal function $\Phi(x)$,
(\ref{5.5}), also transforms homogeneously, because it solves the
homogeneous equation with the operator (\ref{5.18}) and satisfies
the same unit boundary conditions,
    \begin{eqnarray}
      \bar{\Phi}(x)\equiv
      \Phi(x)[\,\bar{g}_{\mu\nu}]
      =\frac{\Phi(x)}{\Psi(x)}.              \label{5.21}
    \end{eqnarray}
One more interesting property is the conformal transformation of
the scalar curvature which reads
    \begin{eqnarray}
     \bar R(x)=-4\,\frac{d-1}{d-2}\,
     \Psi^{-1-4/(d-2)}(x)\,
     \Big[\, F(\nabla)\Psi(x)\,\Big]    \label{5.22}
    \end{eqnarray}
and, thus, means that $\Phi(x)$ can be regarded as a special case
of the conformal transformation, $\Psi=\Phi$, to the conformal
gauge of vanishing scalar curvature, $\bar R=0$.

Consider now the bulk part of the leading asymptotics $W_0$,
(\ref{5.7}), for the conformal operator (\ref{5.18})
    \begin{eqnarray}
     W_0[\,g\,]=-\frac14\,\frac{d-2}{d-1}\,
      \int dx\,g^{1/2} R\,\Phi
      +\frac16\,\Sigma[\,g_\infty].            \label{5.22a}
    \end{eqnarray}
In virtue of (\ref{5.22}) it transforms as
    \begin{eqnarray}
     &&-\frac14\,\frac{d-2}{d-1}\,
    \int dx\,\bar{g}^{1/2}\bar R \,\bar\Phi=
     \int dx\,g^{1/2}\Phi\, F(\nabla)\Psi \nonumber\\
     &&\qquad\qquad\qquad\qquad\qquad\qquad\qquad=
     \int_\infty d\sigma^\mu\,
     (\partial_\mu\Psi-\partial_\mu\Phi).       \label{5.23}
    \end{eqnarray}
Here we integrated by parts, used the equation $F(\nabla)\Phi=0$
and took into account that $\Phi$ and $\Psi$ equal one at
infinity (but their derivatives generically do not tend to zero
fast enough to discard the surface integral). The second term of
the surface integral can be transformed back to the form of the
bulk integral (not involving $\Psi$ now)
    \begin{eqnarray}
    -\int_\infty d\sigma^\mu\,\partial_\mu\Phi
    =-\int dx\,g^{1/2}\,\Box\Phi=
    -\frac14\,\frac{d-2}{d-1}\,
    \int dx\,g^{1/2} R\,\Phi,                  \label{5.24}
    \end{eqnarray}
whence finally
    \begin{eqnarray}
    \int dx\,\bar{g}^{1/2}\bar R \,\bar\Phi=
     \int dx\,g^{1/2} R\,\Phi
     -4\;\frac{d-1}{d-2}\,
    \int_\infty d\sigma^\mu
     \,\partial_\mu\Psi.               \label{5.25}
    \end{eqnarray}
Together with the conformal transformation of the Gibbons-Hawking
integral (\ref{5.8}),
    \begin{eqnarray}
     \bar\Sigma\equiv\Sigma[\,\bar{g}_\infty]=\Sigma
     -4\;\frac{d-1}{d-2}\,
     \int_\infty d\sigma^\mu
     \,\partial_\mu\Psi,             \label{5.26}
    \end{eqnarray}
this means that the following linear combination
    \begin{eqnarray}
     S_\mathrm{C}[\,g\,]=
      -\int dx\,g^{1/2} R\,\Phi+\Sigma[\,g_\infty]  \label{5.27}
    \end{eqnarray}
is conformally invariant in the class of local Weyl
transformations (\ref{5.19})
    \begin{eqnarray}
     S_\mathrm{C}[\,g\,]=S_\mathrm{C}[\,\bar{g}\,].   \label{5.28}
    \end{eqnarray}

This particular combination can be obtained from the Einstein
action (\ref{5.17}) by transition to the conformal gauge of
vanishing scalar curvature. By performing the transformation
(\ref{5.19}) with $\Psi=\Phi[\,g\,]$ one finds that
    \begin{eqnarray}
     S_\mathrm{C}[\,g_{\mu\nu}]=
     S_\mathrm{E}\left[\,g_{\mu\nu}
       \Phi^{4/(d-2)}[\,g\,]\,\right].   \label{5.29}
    \end{eqnarray}
Conformal invariance of this expression is obvious because
$g_{\mu\nu}\Phi^{4/(d-2)}[\,g\,]$ is the invariant of the
conformal transformation (\ref{5.19}). In fact, this functional in
4-dimensional context was suggested in \cite{FradVilk} as a
conformal off-shell extension of the Einstein theory in vacuum (or
with traceless matter sources).

Consider now the conformal properties of the heat kernel
asymptotics for the operator (\ref{5.18}). For the heat kernel
itself they directly follow from the transformation of the
function $\Phi(x)$ (\ref{5.21}) and read in the leading order as
    \begin{eqnarray}
    \bar\Omega(x,y)\,\bar{g}^{1/2}(y)
    =\Psi^{-1}(x)\,\Big[\,\Omega(x,y)\,g^{1/2}(y)\,\Big]\,
    \Psi^{1+4/(d-2)}(y).                  \label{5.30}
    \end{eqnarray}
This transformation looks similar to that of the operator
(\ref{5.20}). However, this coincidence is accidental, because
unlike (\ref{5.20}) the heat kernel does not transform by a simple
homogeneous law. Equation (\ref{5.20}) does not represent the
\emph{similarity} transformation -- the factors on the left and
right of $F(\nabla)$ in (\ref{5.20}) are not inverse to one
another. The extra factor $\Psi^{-4/(d-2)}$ in $\bar
F(\nabla)=\Psi^{-1}\left(\Psi^{-4/(d-2)}\,F(\nabla)\right)\Psi$
indicates that the operator $\bar F(\nabla)$ is similar to the
other operator $\Psi^{-4/(d-2)}\,F(\nabla)$ having another heat
kernel. In fact, this factor, which determines the conformal
rescaling of metric (\ref{5.19}), leads to the exactly calculable
conformal anomaly of the effective action. But for the heat kernel
itself and its functional trace it generates a nontrivial
transformation law. For the functional trace this transformation
with the infinitesimal $\delta\Psi(x)$ in
$\Psi(x)=1+\delta\Psi(x)$ is
    \begin{eqnarray}
    &&\delta_\Psi{\Tr}
    \bar K(s)\Big|_{\Psi=1}=
    -\frac 4{d-2}\,s\frac{d}{ds}\,{\Tr}
    \Big(\delta\Psi\,K(s)\Big)\nonumber\\
    &&\qquad\qquad\qquad\quad
    =-\frac 4{d-2}\,
    s\frac{d}{ds}\,\int dx\,
    \delta\Psi(x)\,\Omega(s\,|\,x,x).          \label{5.32}
    \end{eqnarray}

When $\delta\Psi(x)$ has a compact support we can use here the
leading order asymptotics $\Omega(s\,|\,x,x) \sim\Phi^2(x)/(4\pi
s)^{d/2}$, because in this case the domain of $|x|\to\infty$
(which breaks the uniformity of this asymptotics) does not enter
the range of integration, and $\delta_\Psi{\Tr}K(s)=O(1/s^{d/2})$
at $s\to\infty$. Therefore, the leading asymptotics $W_0$ is
conformal invariant for Weyl rescalings with a compact support.
Indeed, both of the terms in (\ref{5.33}) are separately invariant
in view of (\ref{5.25}) and (\ref{5.26}).

For generic conformal transformations with $\delta\Psi(x)\to 0$
but $\delta\Psi(x)\neq 0$ at $|x|\to\infty$ this conclusion cannot
be inferred on the basis of (\ref{5.32}) because it starts
involving the domain where the non-uniform asymptotics of
$\Omega(s\,|\,x,x)$ breaks down. Therefore, for such
transformations $W_0$ should not necessarily be conformal
invariant, which explains the mismatch of coefficients of the bulk
and surface terms in (\ref{5.33}) as compared to the conformal
extension of the Einstein action (\ref{5.27}).

Quite interestingly, in the exceptional case of the 4-dimensional
spacetime, this invariance holds for $W_0$ even for
transformations with non-compact support. In this case, the
coefficient of nonminimal curvature-scalar coupling
$\,(d\!-\!2)/4(d\!-\!1)=1/6\,$ and
    \begin{eqnarray}
     W_0[\,g\,]=\frac16\,S_\mathrm{C}[\,g\,],\quad d=4.   \label{5.31}
    \end{eqnarray}

\subsection{Problems with the subleading order}
\hspace{\parindent} Unfortunately, inclusion of gravity results in
the number of problems in the subleading order of late time
expansion. The equation for $\Omega_1(x,y)$ generalizing
(\ref{3.13}) takes the form
    \begin{eqnarray}
     F(\nabla)\,\Omega_1(x,y)=
     \left(\sigma^\mu(x,y)\nabla_\mu
     +\frac12\,\Box\,\sigma(x,y)
     -\frac{d}2\,\right)\Phi(x)\,\Phi(y)   \label{5.31a}
    \end{eqnarray}
and has a symmetric solution similar to (\ref{3.14}) \cite{nneag}
  \begin{eqnarray}
  &&\Omega_1(x,y)=\frac12\,\psi(x,y)
  +\frac12\,\psi(y,x)\nonumber\\
  &&\qquad\qquad\qquad\qquad
  -\frac12\,\frac1{F(\nabla_x)}
  \stackrel{\rightarrow}{F}\!(\nabla_x)\,
  [\,\Phi(x)\,\sigma(x,y)\,\Phi(y)\,]
  \stackrel{\leftarrow}{F}\!(\nabla_y)
  \frac{\stackrel{\leftarrow}{1}}{F(\nabla_y)}.   \label{5.32a}
  \end{eqnarray}
Here $\psi(x,y)$ is a special two-point function
    \begin{eqnarray}
    \psi(x,y)=\frac1{F(\nabla_x)}
    \Big[\,2\sigma^\mu(x,y)\nabla_\mu\Phi(x)
     +(\Box\,\sigma(x,y)
     -d)\,\Phi(x)\Big]\,\Phi(y)         \label{5.33}
    \end{eqnarray}
and the arrows indicate the action of the differential operators
in the direction opposite to Green's functions, $1/F(\nabla)$,
written in the operator form. This certainly implies that the
integration by parts that would reverse the action of $F(\nabla)$
on $1/F(\nabla)$ (and, thus, would lead to a complete cancellation
of the corresponding nonlocality) is impossible without generating
nontrivial surface terms.

As shown in \cite{nneag}, with this expression for the
$\Omega_1(x,y)$ the variational equation for $W_1$,
    \begin{eqnarray}
    \frac{\delta \,W_1}{\delta V(x)}=
    -g^{1/2}(x)\,\Omega_1(x,x),         \label{2.13}
    \end{eqnarray}
has the following formal solution in terms of the Green's function
of $F(\nabla)$
    \begin{eqnarray}
    &&W_1[\,V,g_{\mu\nu}]=\frac12\,\int dx\,g^{1/2}(x)\,
    \frac1{F(\nabla_x)}\stackrel{\rightarrow}{F}\!(\nabla_x)\,
    \Big[\,\Phi(x)\,\sigma(x,y)\,\Phi(y)\Big]
    \stackrel{\leftarrow}{F}\!(\nabla_y)\,
    \Big|_{\;y=x}\nonumber\\
    &&\qquad\qquad\qquad\qquad
    +W^\mathrm{metric}_1[\,g_{\mu\nu}].                    \label{5.34}
    \end{eqnarray}
Here $W^\mathrm{metric}_1[\,g_{\mu\nu}]$ is some purely metric
functional -- a functional integration "constant" for equation
(\ref{2.13}). The latter can be determined only from the metric
variational equation (\ref{5.10}). Quite interestingly, this
equation confirms the above expression with a purely constant
metric-independent
$W^\mathrm{metric}_1[\,g_{\mu\nu}]=\mathrm{const}$ (see Appendix
D).

Unfortunately, the validity of the algorithms (\ref{5.32a}),
(\ref{5.33}) and (\ref{5.34}) can be rigorously established only
in flat spacetime. Problem is that the nonlocal function
$\psi(x,y)$ is well (and uniquely) defined only when the
expression in square brackets of (\ref{5.33}) sufficiently rapidly
goes to zero at spacetime infinity. This expression has two terms,
the first of which has a power law falloff $1/|x|^{d-2}$ at
$|x|\to\infty$ in view of the behavior of $\sigma^\mu(x,y)\sim|x|$
and $\nabla_\mu\Phi(x)\sim 1/|x|^{d-1}$. This makes the
contribution of this term (convolution with the kernel of Green's
function) well defined at least in dimensions $d>4$. On the
contrary, the second term is proportional to the deviation of
geodesics $\Box\,\sigma(x,y)-d$ which has the following rather
moderate falloff
    \begin{eqnarray}
      \Box\,\sigma(x,y)-d\sim\frac1{|x|},
      \quad |x|\to\infty.         \label{5.35}
    \end{eqnarray}
Therefore a purely metric contribution to (\ref{2.9}) turns out to
be quadratically divergent in the infrared
    \begin{eqnarray}
     &&\frac1{F(\nabla_x)}\,
     (\Box\,\sigma(x,y)-d)\Phi(x)\,\Phi(y)\nonumber\\
     &&\qquad\qquad\qquad\quad
     =\int dx'\,G(x,x')(\Box\,\sigma(x',y)-d)
     \Phi(x')\,\Phi(y)=\infty.          \label{5.36}
    \end{eqnarray}

Tracing the origin of this difficulty back to the equation
(\ref{5.31a}) we see that the source term in its right hand side
is $O(1/|x|)$, so that the solution $\Omega_1(x,y)\sim|x|$ is not
vanishing at infinity and, therefore, is not uniquely fixed by
Dirichlet boundary conditions. Some principles of fixing this
ambiguity would certainly regularize the integral in the
definition of $\psi(x,y)$ and uniquely specify all quantities in
the subleading order, but this remains the problem for future.

In flat spacetime the geodesic deviation scalar (\ref{5.35}) is
identically vanishing, because $\sigma(x,y)=|x-y|^2/2$,
    $\,\sigma^\mu(x,y)=(x-y)^\mu$,
    $\,\Box\,\sigma(x,y)=d$.
Therefore, the expression for $\psi(x,y)$ becomes well defined.
Correspondingly, in the square brackets of (\ref{5.34}) only one
term containing
$\nabla^\mu_x\nabla^\nu_y\sigma(x,y)=-\delta^{\mu\nu}$ survives
and yields
    \begin{equation}
    F(\nabla_x)F(\nabla_y)
    \Phi(x)\,\sigma(x,y)\,\Phi(y)=
    -4\,\nabla_\mu\Phi(x)\,\nabla^\mu\Phi(y),
    \end{equation}
so that $\Omega_1(x,y)$ and $W_1$ reduce to the expressions
(\ref{3.14}) and the second term of (\ref{3.15}) correspondingly.

There is a conspicuous mismatch between (\ref{3.15}) and the
flat-space limit of (\ref{5.34}) -- the first unit term of
(\ref{3.15}). This term is missing in (\ref{5.34}) and the attempt
to identify it with the constant,
    \begin{equation}
    W_1^\mathrm{metric}=\int dx\,\times 1,  \label{5.34a}
    \end{equation}
contradicts the covariantization procedure in the transition to
curved spacetime (actually confirmed within the covariant
curvature expansion of \cite{CPTII,CPTIII})
    \begin{equation}
     \Tr K(s)=\frac1{(4\pi s)^{d/2}}\int dx\,\times 1+...
     \to\frac1{(4\pi s)^{d/2}}
     \int dx\,g^{1/2}(x)\times 1+...\; .           \label{6.1}
    \end{equation}
The metric-independent integral (\ref{5.34a}) seems
counterintuitive because of its noncovariance. The origin of this
situation may be ascribed to the difficulties of the above type.
The derivation of (\ref{5.34}) is based on the operation with
unregulated divergent integrals which casts serious doubt on its
validity. However, the fact that it formally passes a subtle check
of the metric variation (see Appendix D) suggests that under
certain regularization of these infrared divergences (\ref{5.34}),
(\ref{5.34a}) the algorithm will survive beyond flat spacetime.
The compatibility of this algorithm with the covariant
perturbation expansion of \cite{CPTII,CPTIII} might be nontrivial
in view of these divergences and based on disentangling the
covariant integral $\int dx\,g^{1/2}\times 1$ from divergent
nonlocal structures of (\ref{5.34}), (\ref{5.34a}). This procedure
is discussed in the concluding section below.

The infrared divergence of the integral $\int dx\,g^{1/2}\times 1$
reflects the continuity of the spectrum of the operator
$F(\nabla)$, and in flat spacetime represents a trivial constant
which does not affect physical predictions. In curved spacetime
this integral becomes a functional of geometry and, thus,
incorporates the cosmological constant term. Therefore, the
subleading order of the late-time expansion gets intertwined with
the cosmological constant problem, which we briefly discuss in
Conclusions.

\section{Discussion and conclusions}
\hspace{\parindent} Main results presented above include the
nonperturbative heat kernel asymptotics (\ref{3.10}) and
(\ref{3.12}), their generalization to curved spacetime
(\ref{5.7})-(\ref{5.8}), and as a byproduct new nonlocal effective
action algorithms (\ref{4.8}), (\ref{4.8a}) and (\ref{4.11}). Let
us now briefly discuss potential generalizations and applications
of these results and the remaining loopholes in our technique.

Nonperturbative effective action alternative to the
Coleman-Weinberg potential is very interesting in various
applications. Already at this preliminary stage it is clear that
the algorithms (\ref{4.8}) and (\ref{4.8a}) show nontrivial and
qualitatively important dependence on spacetime dimensionality,
because in $d>4$ the action is dominated by the
\emph{renormalization-independent} structure not involving the
parameter $\mu^2$. This is very different from the
four-dimensional case (\ref{4.8}) when the infrared effects simply
``delocalize" the Coleman-Weinberg potential, but leave the
dominant dependence on the ultraviolet scale parameter. This means
that scaling arguments cannot grasp this infrared dominant term at
all. It is also interesting that this term is
\emph{negative-definite}, which might have important implications.
At the same time, the nonperturbative terms deserve further
analysis ascertaining their range of applicability, the
corrections due to local derivative factors completely disregarded
above (terms with $\tilde{a}_n(x,x)$ for $n>0$).

Regarding the general formalism of late time expansion it should
be emphasized that the difficulty we encountered in the subleading
order indicates a serious problem. This difficulty will
proliferate in higher orders of the $1/s$-expansion, for which the
coefficients $\Omega_n(x,y)$ at spacetime infinity will behave as
higher and higher powers in $|x|$ and $|y|$ and thus comprise
increasingly more complicated boundary value problems. This will
generally happen not only in metric sector, but in flat spacetime
as well.

The subleading order is free from this problem in flat spacetime,
because the expressions (\ref{3.14}) and (\ref{3.15}) are well
defined, but inclusion of gravity results in the problem discussed
above. However, already this order of $1/s$-expansion is very
interesting, because it incorporates the cosmological constant
problem. Indeed, the cosmological term is generated via the
integral (\ref{1.10}) from the (covariantized) unit term in
${\Tr}K(s)$, (\ref{6.1}), as
    \begin{eqnarray}
     \SGamma_\Lambda=\int dx\,g^{1/2}\,
     \Lambda_\infty,\quad
     \Lambda_\infty=
     -\frac1{2(4\pi)^{d/2}}\int_0^\infty
     \frac{ds}{s^{1+d/2}}\times 1.                   \label{6.2}
    \end{eqnarray}
The cosmological constant $\Lambda_\infty$ here is ultraviolet
divergent, and this expression is also infrared divergent in the
coordinate sense\footnote{Of course, the variety of divergences
indicates that the cosmological constant cannot consistently arise
in asymptotically-flat spacetime. The contribution (\ref{6.2}) in
\emph{massless} theories does not carry any sensible physical
information and is cancelled due to a number of interrelated
mechanisms. Its cancellation is guaranteed by the contribution of
the local path-integral measure to the effective action, which
annihilates strongest (volume) divergences \cite{Bern}. Another
mechanism is based on the use of the dimensional regularization
which puts to zero all power-like divergences. All these
mechanisms, however, stop working for \emph{massive} theories or
for theories with spontaneously broken symmetry, where the induced
vacuum energy presents a real hierarchy problem \cite{Weinberg}.}
-- the volume integral $\int dx\,g^{1/2}$ for asymptotically-flat
spacetime diverges at $|x|\to\infty$.

Therefore, the covariance problem for $W_1$ with
metric-independent $W_1^\mathrm{metric}$, given by Eqs.
(\ref{5.34}) and (\ref{5.34a}), amounts to correctly recovering
the covariant cosmological term from the nonlocal (and divergent)
expressions. To make the discussion of this point simpler,
consider a purely metric case of vanishing potential, $V=0$,
$\Phi=1$. For this case the subleading term of the functional
trace (with all the reservations discussed in Sect.5.3) is
anticipated to be
    \begin{eqnarray}
    W_1=\int dx\times 1+\frac12\,\int
    dx\,dy\,g^{1/2}(x)\,G(x,y)\,
    \Box_x\Box_y\sigma(x,y).               \label{6.3}
    \end{eqnarray}
The second integral obviously vanishes in flat spacetime where
$\Box_x\Box_y\sigma(x,y)=0$, so that its curvature expansion
should begin with the first order. This integral is infrared
divergent and, moreover, as it follows from calculations of
Appendix D, it is quadratically divergent, just like the first
order term in the metric (or curvature) expansion of
    \begin{eqnarray}
    \int dx\,g^{1/2}(x)=
    \int dx\left(1+\frac12\, h+...\right)
    =\int dx\left(1-\frac1\Box R(x)+...\right).  \label{6.4}
    \end{eqnarray}
Comparison with (\ref{6.3}) suggests that the second integral of
(\ref{6.3}) begins with this nonlocal term linear in $(1/\Box)R$
(or local in $h$). Thus the total cosmological term seems to be
camouflaged in (\ref{6.3}) as a sum of a trivial local and a
special nonlocal functionals\footnote{The idea of nonlocal
representation of the cosmological term was also considered in
\cite{nonlocal,shap}.},
    \begin{equation}
    W_1=\int dx\,g^{1/2}\times 1+O\Big(\,R^2\,\Big).  \label{6.5}
    \end{equation}

This situation is not entirely new. We have already seen that the
local surface integral of Gibbons and Hawking has a nonlocal
representation (\ref{5.15}) which, in its turn, underlies the
nonlocal representation of the Einstein action of \cite{nonlocal}.
This analogy is, however, marred by the fact that the
corresponding integrals in (\ref{6.3}) and (\ref{6.4}) are
infrared divergent, and special regularization is needed to make
their rigorous comparison. This regularization will apparently be
a part of the scheme necessary for rendering the subleading order
of the $1/s$-asymptotics (and maybe all of its higher orders)
well-defined. The invention of this regularization will result in
two possible outcomes -- it will either confirm the equation
(\ref{6.5}) or show that the cosmological term $\int dx\,g^{1/2}$
enters ${\Tr}K(s)$ at $s\to 0$ and $s\to\infty$ with different
coefficient functions of $s$.

The second option seems rather unlikely, but if it happens, this
would mean a new mechanism of induced cosmological constant.
Indeed, in this case the $s$-dependent (or partial) "cosmological
constant" $\lambda(s)$ in ${\Tr}K(s)$, $\,{\Tr}K(s)=\lambda(s)\int
dx\,g^{1/2}+...$, would interpolate between $\lambda_<(s)=1/(4\pi
s)^{d/2}$ at $s\to 0$ and some other function $\lambda_>(s)$ at
$s\to\infty$. This would result in the full induced cosmological
constant
     \begin{eqnarray}
     \Lambda_\mathrm{ind}=-\frac12\int_0^\infty
     \frac{ds}s\,\lambda(s)\times 1\neq 0,                   \label{6.6}
     \end{eqnarray}
that will definitely be non-vanishing, because in contrast to
(\ref{6.2}) this integral no longer represents a pure power
divergence annihilated by the dimensional
regularization\footnote{This interesting mechanism will require
additional dimensionful parameter, and maybe the absence of such
parameter will preclude this mechanism from realization.}. This
possibility is currently under study. We expect that this might
bring to light interesting interplay between the cosmological
constant problem and infrared asymptotics of the heat kernel and
nonlocal effective action.

Let us conclude by listing possible generalizations of our late
time technique. Once the problems with subleading order are
resolved, one can extend the effective action algorithms
(\ref{4.8}) and (\ref{4.8a}) to include curvature. This extension
is interesting, because it might provide us with long-distance
modifications of gravity theory characterized by the nonlocal
scale-dependent gravitational "constant" \cite{nonlocal,A-HDDG}.
Another generalization consists in overstepping the limits of the
asymptotically-flat spacetime. The first thing to do here is to
consider asymptotically deSitter boundary conditions which are
strongly motivated by the cosmological acceleration phenomenon and
by the dS/CFT-correspondence conjecture \cite{dS/CFT}. This
generalization implies essential modification of both perturbative
and nonperturbative techniques for the heat kernel, the
generalization of the Gibbons-Hawking term to asymptotically
dS-spacetimes, etc. Another generalization concerns the inclusion
of higher spins with the covariant derivatives in the
d'Alembertian involving not only the metric connection but the
gauge field connection as well. All these issues are currently
under study and will be presented elsewhere.

%\vskip 1cm
\section*{Acknowledgements}

A.O.B. is grateful for hospitality of the Erwin Schroedinger
Institute for Mathematical Physics and thanks the participants of
the Workshop "Gravity in Two Dimensions" for helpful stimulating
discussions. This work was supported by the RFBR grant No
02-01-00930 and the LSS grant No 1578.2003.2. The work of D.V.N.
was also supported by the RFBR grant No 02-02-17054 and by the
Landau Foundation.

\vskip 1cm

\setcounter{section}{0}
\renewcommand{\theequation}{\Alph{section}.\arabic{equation}}
\renewcommand{\thesection}{\Alph{section}.}
\renewcommand{\thesubsection}{\Alph{section}.\arabic{subsection}.}

\section{Late time asymptotics in perturbation theory}
\hspace{\parindent} Substituting the expression (\ref{3.19}) for
$\Omega _{n}$ in (\ref{B1}) and expanding in powers of the term
bilinear in $\alpha$-parameters one gets
    \begin{eqnarray}
    &&{\Tr}K_{n}(s)=
    \frac{(-s)^{n}}{(4\pi s)^{d/2}}\int
    dx\,\int_{0}^{\infty }d^{n-1}\alpha \,
    \exp \Big(s\sum_{i=2}^{n}\alpha_{i}D_{i}^{2}\Big)  \nonumber \\
    &&\qquad \qquad \qquad \qquad \qquad
    \times \left( 1-s\sum_{m,k=2}^{n}\alpha_{m}
    \alpha _{k}D_{m}D_{k}+...\right) \,V_{1}V_{2}...V_{n}.
    \end{eqnarray}
Here $1/n$ factor disappeared due to the contribution of $n$ equal
terms corresponding to the operatorial maxima of $\Omega _{n}$.
Also the range of integration over $\alpha_{2},...\alpha_{n}$,
$\sum_{i=2}^{n}\alpha _{i}\leq 1$, was extended to all positive
values of $\alpha _{i}$. This is justified since the error we make
by this extension goes to higher orders of
$1/s$-expansion\footnote{If $\Omega _{n}$ were not a differential
operator this error would be exponentially small in $s\to\infty$.
Because of the heat-kernel operator nature, however, it turns out
to be suppressed by extra power-like factor $O(1/s^{d/2})$ and,
therefore, goes beyond the leading and subleading orders.}. The
second term in the round brackets can be rewritten in terms of the
derivatives with respect to $D_{m}^{2}$ acting on the exponential,
so that
    \begin{eqnarray}
    &&{\Tr}K_{n}(s)=\frac{(-s)^{n}}{(4\pi s)^{d/2}}
    \int dx\,\left( 1-\frac{1}{s}\sum_{m,k=2}^{n}D_{m}D_{k}
    \frac{\partial }{\partial D_{m}^{2}}\frac{
    \partial }{\partial D_{k}^{2}}+...\right)   \nonumber \\
    &&\qquad \qquad \qquad \qquad \qquad \times
    \int_{0}^{\infty }d^{n-1}\alpha
    \,\exp \Big(s\sum_{i=2}^{n}
    \alpha _{i}D_{i}^{2}\Big)\,V_{1}V_{2}...V_{n}.
    \end{eqnarray}
In this form it is obvious that further terms of expansion in
powers of the quadratic in $\alpha $ part of $\Omega _{n}$ bring
higher order corrections of the $1/s$-series. Doing the integral
over $\alpha$ here and performing differentiations one obtains
    \begin{eqnarray}
    &&{\Tr}K_{n}(s)=\frac{1}{(4\pi s)^{d/2}}\int dx\,
    \left[ -s\frac{1}{D_{2}^{2}...D_{n}^{2}}
    +2\sum_{m=2}^{n}\frac{1}{D_{2}^{2}...D_{m-1}^{2}}\,
    \frac{1}{(D_{m}^{2})^{2}}\,
    \frac{1}{D_{m+1}^{2}...D_{n}^{2}}\right. \nonumber \\
    &&\qquad +2\sum_{m=2}^{n-1}\sum_{k=m+1}^{n}
    \frac{1}{D_{2}^{2}...D_{m-1}^{2}}
    \,\frac{D_{m}^{\mu }}{(D_{m}^{2})^{2}}\,
    \frac{1}{D_{m+1}^{2}...D_{k-1}^{2}}\,
    \frac{D_{k\mu }}{(D_{k}^{2})^{2}}\,
    \frac{1}{D_{k+1}^{2}...D_{n}^{2}}     \nonumber \\
    &&\qquad \qquad \qquad \qquad \qquad \qquad
    \left. +O\left( \frac{1}{s}
    \right) \right] \,V_{1}V_{2}...V_{n}.  \label{bigequation}
    \end{eqnarray}

The first term in the square brackets gives the leading order term
of the late time expansion. It can be further transformed by
taking into account that any operator $D_{m}$ defined by (\ref{D})
acts as a partial derivative only on the group of factors
$V_{m}V_{m+1}...V_{n}$ in the full product $V_{1}...V_{n}$,
$D_{m}V_{1}...V_{n}=V_{1}...V_{m-1}\nabla (V_{m+1}...V_{n})$.
Therefore all the operators understood as \emph{acting to the
right} can be ordered in such a way
    \begin{equation}
    {\Tr}K_{n}(s)=-\frac{s}{(4\pi s)^{d/2}}
    \int dx\,V_{1}\frac{1}{D_{n}^{2}}
    V_{n}\frac{1}{D_{n-1}^{2}}V_{n-1}...
    \frac{1}{D_{2}^{2}}V_{2}+O\left( \frac{1}{s^{d/2}}\right)
    \end{equation}
that the labels of $D_{m}^{2}$'s can be omitted and all
$D_{m}^{2}$ can be identified with boxes also acting to the right
    \begin{equation}
    {\Tr}K_{n}(s)=-\frac{s}{(4\pi s)^{d/2}}
    \int dx\,\underbrace{V\frac{1}{\Box }V\frac{1}{\Box }...
    V\frac{1}{\Box }}_{n-1}\,V(x)
    +O\left( \frac{1}{s^{d/2}}\right) .  \label{b1}
    \end{equation}
Infinite summation of this series is not difficult to perform
because this is the geometric progression in powers of the
nonlocal operator $V(1/\Box )$
    \begin{equation}
    {\Tr}K(s)={\Tr}K_{0}(s)
    -\frac{s}{(4\pi s)^{d/2}}\int
    dx\,\sum_{n=0}^{\infty }
    \left( V\frac{1}{\Box }\right) ^{n}V(x)
    +O\left(\frac{1}{s^{d/2}}\right),
    \end{equation}
which gives rise to the leading order in (\ref{3.18}).

The subleading in $s$ terms are given by infinite resummation over
$n$ of the second and third terms in square brackets of
Eq.(\ref{bigequation}). Remarkably, this summation can again be
explicitly done. In this case one has to sum multiple geometric
progressions.

The second term of (\ref{bigequation}) gives rise to the series
    \begin{equation}
    \frac{2}{(4\pi s)^{d/2}}\int dx\,
    \sum_{n=2}^{\infty }\sum_{m=2}^{n}V
    \underbrace{\frac{1}{\Box }V...
    \frac{1}{\Box }V}_{n-m}\frac{1}{\Box ^{2}}
    \underbrace{V\frac{1}{\Box }...
    V\frac{1}{\Box }}_{m-2}\,V(x).                  \label{1}
    \end{equation}
By summing the two geometric progressions with respect to
independent summation indices $0\leq n-m<\infty $ and $0\leq
m-2<\infty$ one finds that this series reduces to
    \begin{equation}
    \frac{2}{(4\pi s)^{d/2}}\int dx\,V\,
    \frac{1}{(\Box -V)^{2}}\,V(x),                    \label{2}
    \end{equation}
which after the integration by parts amounts to
    \begin{equation}
    \frac{2}{(4\pi s)^{d/2}}\int dx\,
    \left( \frac{1}{\Box -V}\,V(x)\right) ^{2}=
    \frac{2}{(4\pi s)^{d/2}}\int dx\,
    \Big(1-\Phi (x)\Big)^{2}.                  \label{B1000}
    \end{equation}

Similarly, the third term of (\ref{bigequation}) gives rise to the
triplicate geometric progression which after summation and
integration by parts reduces to
    \begin{eqnarray}
    &&\frac{2}{(4\pi s)^{d/2}}\int dx\,
    V\sum_{i=0}^{\infty }\Big(\frac{1}{\Box }V
    \Big)^{i}\frac{1}{\Box }\nabla ^{\mu }
    \frac{1}{\Box }\,\sum_{j=0}^{\infty }
    \Big(V\frac{1}{\Box }\Big)^{j}
    \frac{1}{\Box }\nabla ^{\mu }\frac{1}{\Box}
    \sum_{l=0}^{\infty }
    \Big(V\frac{1}{\Box }\Big)^{l}\,V(x)  \nonumber \\
    &&\qquad \qquad \qquad \qquad
    =-\frac{2}{(4\pi s)^{d/2}}\int dx\,
    \Big(\nabla_{\mu }\Phi (x)\Big)
    \frac{1}{\Box -V}V\frac{1}{\Box }
    \nabla ^{\mu }\Phi (x).                   \label{B100}
    \end{eqnarray}
Taking here into account that
    \begin{equation}
    \frac{1}{\Box -V}V\frac{1}{\Box }
    =\frac{1}{\Box -V}-\frac{1}{\Box }
    \end{equation}
one finds that the sum of (\ref{B1000}) and (\ref{B100}) is equal
to
    \begin{eqnarray}
    &&\frac{2}{(4\pi s)^{d/2}}\int dx\,
    \left( (1-\Phi )^{2}-\nabla _{\mu }\Phi
    \frac{1}{\Box -V}\nabla ^{\mu }\Phi
    +\nabla _{\mu }\Phi \frac{1}{\Box }
    \nabla ^{\mu }\Phi \right)   \nonumber \\
    &&\qquad \qquad \qquad \qquad \qquad \qquad
    =-\frac{2}{(4\pi s)^{d/2}}\int
    dx\,\nabla _{\mu }\Phi
    \frac{1}{\Box -V}\nabla ^{\mu }\Phi ,  \label{B10000}
    \end{eqnarray}
where the cancellation of the first and the third terms takes
place after rewriting $\nabla _{\mu }\Phi $ in the third term as
$\nabla _{\mu }(\Phi -1)$ and integrating it by
parts\footnote{Straightforward integration by parts of
$\nabla_{\mu }\Phi (1/\Box )\nabla^{\mu }\Phi $ is impossible
because $\Phi (x)$ does not vanish at $|x|\rightarrow \infty$,
while $\Phi (x)-1$ does.}. This finally recovers the subleading
term of (\ref{3.22}).

\section{Nonperturbative effective action}
\hspace{\parindent} To generate effective action by piecewise
smooth approximation (\ref{4.1})-(\ref{4.5}) we first note that
the equation (\ref{4.1}) for $s_{\ast}$,
    \begin{equation}
    \int dx\,\exp \left( -Vs_{\ast }\right)
    =\int dx\,(1-s_{\ast }V\Phi ),          \label{eqfors}
    \end{equation}
is not exactly solvable. Its solution, as some nontrivial
functional of the potential, $s_{\ast }=s_{\ast }[\,V(x)\,]$, can
however be approximately obtained for two wide classes of
potentials (\ref{4.6}) satisfying the bounds (\ref{4.8}) and
(\ref{4.10}) respectively. As we will now see these bounds also
provide the efficiency of this approximation.

For this purpose we rewrite the action (\ref{4.5}) as a sum of two
contributions
    \begin{equation}
    \bar{\SGamma}=\SGamma_{<}+\SGamma_{>}=
    -\frac{1}{2}\int\limits_{0}^{\infty }
    \frac{ds}{s}{\Tr}K_{<}(s)
    -\frac{1}{2}\int\limits_{s_{\ast }}^{\infty }
    \frac{ds}{s}\left( {\Tr}K_{>}(s)
    -{\Tr}K_{<}(s)\right) .                   \label{ll}
    \end{equation}
The first integral here represents the calculation of Sect.2.2
within the modified Schwinger-DeWitt expansion with
$\tilde{a}_{0}=1$ and $\tilde{a}_{n}=0,\,n\geq 1$, which in our
particular case gives rise to
    \begin{eqnarray}
    &&\SGamma_{<}=\SGamma_\mathrm{CW},             \label{uv}
    \end{eqnarray}
where $\SGamma_\mathrm{CW}$ is just the Coleman-Weinberg potential
(\ref{2.11}) and we disregard the divergent part and all finite
terms reflecting renormalization ambiguity. The second integral in
(\ref{ll}) can also be calculated exactly. Integrating by parts
and taking into account (\ref{eqfors}), we obtain
    \begin{eqnarray}
    &&\SGamma_{>}\equiv
    -\frac{1}{2}\int\limits_{s_{\ast }}^{\infty }
    \frac{ds}{s}\,\left( {\Tr}K_{>}(s)
    -{\Tr}K_{<}(s)\right)       \nonumber \\
    &&\qquad
    =\frac{1}{2 (4\pi s_*)^{d/2}}\int dx\,
    \Big[ \,\frac{4\,s_*V\Phi}{d\,(d-2)}
    +\sum\limits_{n=1}^{d/2-1}\frac{(d/2-n-1)!}{(d/2)!}\,(-s_*V)^n\,
    e^{-s_{\ast}V}\nonumber\\
    &&\qquad \qquad\qquad\qquad\qquad\qquad
    +\frac{(-s_*V)^{d/2}}{(d/2)!}\Gamma (0,s_{\ast}V)\,\Big],  \label{ir}
    \end{eqnarray}
where $\Gamma (0,x)$ is an incomplete gamma function, $\Gamma
(0,x)=\int_{x}^{\infty }\,dt\,t^{-1}e^{-t}$, with the following
asymptotics
    \begin{equation}
    \Gamma (0,x)\sim \left\{
    \begin{array}{lc}
    \ln \displaystyle{\frac{1}{x}}\, , &\quad x\ll 1, \\
    \displaystyle{\frac1x}\,e^{-x}\, , &\quad x\gg 1.
    \end{array}\right.                   \label{gass}
    \end{equation}

Further steps depend on the class of potentials, for which the
consistency of the piecewise approximation should be carefully
analyzed.

\subsection{Small potential}
\hspace{\parindent} The approximation (\ref{4.1}) - (\ref{4.5}) is
efficient only if the ranges of validity of two asymptotics
(respectively for small and big $s$) overlap with each other and
the point $s_{\ast}$ belongs to this overlap. Below we show that
this requirement is satisfied for two classes of potentials
satisfying the bounds (\ref{4.8}) and (\ref{4.10}).

The modified gradient expansion is applicable in this overlap
range of $s$ if
    \begin{equation}
    s\nabla \nabla V\ll V,  \label{B.1}
    \end{equation}
(cf. Eq.(\ref{parexp}) with $s$ replaced by effective cutoff
$s=1/V$). The applicability of the large $s$ expansion in the same
domain reads as
    \begin{equation}
    s\int dx\,V\Phi \gg \int dx\,
    \nabla _{\mu }\Phi \frac{1}{V-\Box }
    \nabla^\mu\Phi,                                  \label{B.4}
    \end{equation}
which means that the subleading term of the late time expansion
(\ref{3.22}) (quadratic in $\nabla_\mu\Phi$) is much smaller than
the leading order term.

To implement these requirements we assume that $V(x)$ has a
compact support of finite size $R$, (\ref{4.6}), and its
derivatives are bounded and satisfy the following estimate
    %\begin{equation}
    $\nabla \nabla V\sim V_{0}/R^{2}$,
    %\end{equation}
so that (\ref{B.1}) reads as $sV_{0}/R^{2}\ll V_{0}$, or
    \begin{equation}
    s\ll R^{2}.                      \label{a2}
    \end{equation}
To find out what does the criterion (\ref{B.4}) mean let us make a
further assumption, namely, that the potential $V$ is small. In
this case it can be disregarded in the Green's functions and
$1/(\Box-V)$ can be replaced by $1/\Box $. Therefore the following
estimates hold
    \begin{eqnarray}
    &&\frac{1}{\Box -V}V(x)\sim \int_{|y|\leq R}dy\,
    \frac{1}{|x-y|^{d-2}}%
    V(y)\sim \frac{1}{R^{d-2}}R^{d}V_{0}
    \sim V_{0}R^{2},                  \nonumber\\
    &&\int dx\,V\Phi \sim V_{0}R^{d}, \nonumber\\
    &&\int dx\,\nabla_\mu\Phi
    \frac{1}{V-\Box }\nabla ^{\mu }\Phi
    \simeq V_{0}^{2}R^{d+4}.                             \label{est}
    \end{eqnarray}
Roughly, every Green's function gives the factor $R^{2}$, every
derivative -- $1/R$, integration gives the volume of compact
support $R^{d}$, etc. Applying these estimates to eq. (\ref{B.4})
we get $sV_{0}R^{d}\gg V_{0}^{2}R^{d+4}$, whence $s\gg
V_{0}R^{4}$. Combining this with (\ref{a2}) one gets the following
range of overlap of our asymptotic expansions
    \begin{equation}
    R^{2}\gg s\gg V_{0}R^{4},  \label{a5}
    \end{equation}
whence it follows that this overlap domain is not empty only if
    \begin{equation}
    V_{0}R^{2}\ll 1.  \label{sp}
    \end{equation}
Moreover, the assumption of disregarding the potential in the
Green's function is also justified in this case since $V\sim
V_{0}\ll 1/R^{2}\sim \Box $.

Now let us check whether $s_{\ast }$ introduced above belongs to
the overlap domain (\ref{a5}). Note that if it is really so, then
$s_{\ast}V$ in Eq.(\ref{eqfors}) is much smaller than unity
because in the overlap range one has $sV\sim sV_0\ll R^2 V_0\ll
1$. Hence the exponent in the left hand side of (\ref{eqfors}) can
be expanded in powers of $s_{\ast }V$, and the resulting equation
for $s_{\ast}$ becomes\footnote{Note that the quadratic term
should be retained in the expansion of $e^{-s_*V}$ if we want to
get a nontrivial solution for $s_{\ast}$.}
    \begin{equation}
    \int dx\,\left( 1-s_{\ast}V
    +\frac{s_{\ast }^{2}}{2}V^{2}
    +O\left( \left(s_{\ast}V\right)^{3}\right) \right)
    =\int dx\,\left( 1-s_{\ast}V\Phi\right).     \label{B.5}
    \end{equation}
Its solution has the following form:
    \begin{equation}
    s_{\ast}\simeq 2\,
    \frac{\vphantom{\displaystyle\frac{a}a}\int dx\,
    V(1-\Phi )}{\vphantom{\displaystyle\frac{a}a}\int dx\,V^{2}}=
    2\,\frac{\vphantom{\displaystyle\frac{a}a}\int dx\,
    V\!\frac{1}{V-\Box }V}
    {\vphantom{\displaystyle
    \frac{a}a}\int dx\,V^{2}}.          \label{B.6}
    \end{equation}
Taking into account the estimates (\ref{est}) we see that the
point $s_{\ast}\sim R^{2}$ belongs to the upper edge of the
interval (\ref{a5}). Late time expansions is fairly well satisfied
here, but the small $s$ expansion is on the verge of breakdown. At
this level of generality it is hard to overstep the uncertainty of
this estimate. There is a hope that numerical coefficients in more
precise estimates (with concrete potentials) can be large enough
to shift $s_{\ast}$ to the interior of the interval (\ref{a5})
and, thus, make our approximation completely reliable.

Bearing in mind all these reservations let us proceed with the
calculation of the effective action. We have to use the small
$x=s_*V$ asymptotics (\ref{gass}) in the expression (\ref{ir}).
This leaves us with the first two dominant terms in the right hand
side of (\ref{ir}) plus the logarithmic term coming from the
incomplete gamma function.  Then we apply Eq.(\ref{B.6}) to
express the integral of $V(1-\Phi)$ in terms of $s_*$ and that of
$V^2$ and get
    \begin{equation}
    \SGamma_{>}\simeq \frac{1}{2\,(4\pi)^{d/2}}
    \int dx\,\left[ \,-\frac{2}{d\,(d-2)}\frac{V^2}{s_*^{d/2-2}}
    +\frac{(-V)^{d/2}}{(d/2)!}\,
    \ln \frac{1}{s_*\mu^2}\,\right] -\SGamma_\mathrm{CW}, \label{B.7}
    \end{equation}
where the $d$-dimensional Coleman-Weinberg term
$\SGamma_\mathrm{CW}=\SGamma_\mathrm{CW}(\mu^2)$, (\ref{2.11}),
was disentangled from the logarithm of the incomplete gamma
function. Therefore, in the whole action
$\bar{\SGamma}=\SGamma_{<}+\SGamma_{>}$ the Coleman-Weinberg term
gets cancelled and the final answer reads as (\ref{4.8a}).

For generic potential satisfying the smallness bound (\ref{4.7})
$s_*V\ll 1$, so that the first term in square brackets of
(\ref{B.7}) is much bigger than the logarithmic one,
$V^2/s_*^{d/2-2}\gg V^{d/2}$ (confer Eq.(\ref{4.8b})). However, in
four dimensions, $d=4$, both terms become of the same order of
magnitude and the first term takes the form of the finite
counterterm $\sim \int d^4x\,V^2$ to ultraviolet divergences,
which we disregard (or absorb in the redefinition of the
$\mu^2$-parameter). Therefore, in the four-dimensional case only
the logarithmic term survives and gives rise to (\ref{4.8}).

\subsection{Big potential}

\hspace{\parindent} Remarkably, the case of the small potentials
(\ref{4.7}) is not the only one when one can find a non-empty
domain of overlap where both asymptotics for ${\Tr}K(s)$ are
applicable. Namely, the opposite case of big potentials (in units
of the inverse size of their support), (\ref{4.10}), is equally
good. The key observation here is that in this case the kernel of
the Green's function $1/(\Box -V)$ can be replaced within the
compact support of $V$ by $-1/V$ ($\Box \sim 1/R^{2}\ll V_{0}\sim
V$) and correspondingly
    \begin{eqnarray}
    &&\frac{1}{\Box -V}V(x)
    \sim -\frac{1}{V}V=-1,  \label{B.12a} \\
    &&\int dx\,\nabla _{\mu }
    \Phi \frac{1}{V-\Box }
    \nabla ^{\mu }\Phi \simeq
    \frac{R^{d}}{V_{0}R^{2}}.  \label{B.12}
    \end{eqnarray}
Therefore, the criterion of applicability of the late time
expansion (\ref {B.4}) becomes $s\gg 1/V_{0}^{2}R^{2}$. Together
with (\ref{a2}) it yields the new overlap range
    \begin{equation}
    R^{2}\gg s\gg \frac{1}{V_{0}^{2}R^{2}}  \label{bpint}
    \end{equation}
which is obviously not empty if the potential satisfies
(\ref{4.10}).

To find $s_{\ast}$ in this case we have to solve the equation
(\ref{eqfors}) for the case when $s_{\ast }V$ is not anymore a
small quantity. Since $V$ is big, the exponent in (\ref{eqfors})
can be replaced by zero inside the compact support,
$\exp(-s_{\ast}V(x))\sim 0$, $|x|\leq R$, and by one outside of it
where the potential vanishes, $\exp (-s_{\ast }V(x))\sim 1$,
$|x|>R$. Rewriting the integrals in both sides of the equation
(\ref{eqfors}) as a sum of contributions of $|x|\leq R$ and
$|x|>R$, we see that the contribution of the non-compact domain
gets cancelled and the equation becomes
    \begin{equation}
    s_{\ast }\int\limits_{|x|\leq R}dx\,
    V\Phi \simeq \int\limits_{|x|\leq R}dx.   \label{B.12b}
    \end{equation}
Then it follows that $s_{\ast}$ is approximately given by the
inverse of the function $V\Phi (x)$ \emph{averaged} over the
compact support of the potential
    \begin{eqnarray}
    s_{\ast }\simeq \frac{1}{\big<V\Phi \big>},
    \end{eqnarray}
where $\big<V\Phi \big>$ is given by (\ref{4.12}).

A qualitative estimate of $\big<V\Phi \big>\sim V_{0}$ implies
that $s_{\ast}\sim 1/V_{0}$ and it belongs to the middle of the
interval (\ref{bpint}). This makes the case of big potentials
fairly consistent. On the other hand, the value of $\Phi (x)$ is
close to zero inside the potential support (see (\ref{B.12a})), so
most likely the estimate for $\big<V\Phi \big>$ is smaller by at
least one power of the quantity $1/V_{0}R^{2}$, which is the basic
dimensionless small parameter in this case. Therefore the
magnitude of $s_{\ast}$ becomes bigger by one power of
$V_{0}R^{2}$, $s_{\ast }\simeq R^{2}$, which is again near the
upper boundary of the overlap interval (\ref{bpint}). Similarly to
the small potential case, a more rigorous analysis is needed
(maybe for more concretely specified potentials) to account for
subtle edge effects at the boundary of compact support, which
might shift the value of $s_{\ast}$ to a safe region inside
(\ref{bpint}).

With the above estimate for $s_{\ast}\sim R^{2}$ the magnitude of
$s_{\ast}V$ in the expression for the infrared part of the
effective action (\ref{ir}) becomes big, $s_{\ast}V\sim
s_{\ast}V_{0}\sim V_{0}R^{2}\gg 1$. Therefore, we use the big $x$
asymptotics (\ref{gass}) in (\ref{ir}) and get on account of
(\ref{B.12b}) the contribution
    \begin{equation}
    \SGamma_{>}\simeq \frac{1}{(4\pi ^{2}s_{\ast
    })^{d/2}}\,\frac{2s_\ast}{d\,(d-2)}
    \int\limits_{|x|\leq R} dx\,V\Phi =\frac{1}{(4\pi)^{d/2}}\,
    \frac{2\,\big<V\Phi \big>^{d/2}}{d\,(d-2)}\,\int\limits_{|x|\leq R}
    \!dx\,.
    \end{equation}
In this case the Coleman-Weinberg term is not cancelled in
complete agreement with what we would expect for big potentials
and the final result reads as (\ref{4.11}).

\section{Nonlocal form of the Gibbons-Hawking surface integral}
\hspace{\parindent} To check the relation (\ref{5.15}) we first
calculate the metric variational derivative of its left hand side
at finite $|x|$ and show that it is zero with the needed accuracy
in powers of the curvature. The variations of the linear,
quadratic and cubic in curvature terms give respectively
   \begin{eqnarray}
    &&\frac{\delta}{\delta g_{\mu\nu}}
    \int dx\,g^{1/2}\,R
    =-g^{1/2}\,\Big(R^{\mu\nu}
    -\frac12g^{\mu\nu}R\Big),                      \label{GH.5.1}\\
    &&\frac{\delta}{\delta g_{\mu\nu}}
    \int dx\,g^{1/2}\,
    \left\{-R_{\mu\nu}\frac1{\Box} R^{\mu\nu}
    +\frac12\,R\frac1{\Box} R\right\}
    \nonumber\\
    &&\qquad\qquad\qquad\qquad\qquad\quad=
    g^{1/2}\,\Big(R^{\mu\nu}
    -\frac12g^{\mu\nu}R\Big)
    +g^{1/2}\,J^{\mu\nu}
    +\mathrm{O}[\,R_{\mu\nu}^3\,],                 \label{GH.5.2}\\
    &&\frac{\delta}{\delta g_{\mu\nu}}
    \int dx\,g^{1/2}\,\left\{\,
    \frac12\,R\left(\frac1{\Box}
    R^{\mu\nu}\right)\frac1{\Box} R_{\mu\nu}
    -R^{\mu\nu}\left(\frac1{\Box}
    R_{\mu\nu}\right)\frac1{\Box} R\right.
    \nonumber\\
    &&\qquad\qquad
    +\left(\frac1{\Box} R^{\alpha\beta}\right)
    \left(\nabla_\alpha\frac1{\Box} R\right)
    \nabla_\beta\frac1{\Box} R
    \nonumber\\
    &&\qquad\qquad
    -2\,\left(\nabla^\mu\frac1{\Box} R^{\nu\alpha}\right)
    \left(\nabla_\nu\frac1{\Box}
    R_{\mu\alpha}\right)\frac1{\Box} R
    \nonumber\\
    &&\qquad\qquad
    -\left.2\,\left(\frac1{\Box} R^{\mu\nu}\right)
    \left(\nabla_\mu\frac1{\Box}
    R^{\alpha\beta}\right)\nabla_\nu\frac1{\Box}
    R_{\alpha\beta}\,\right\}
 %   \nonumber\\
 %   &&\qquad\qquad\qquad\qquad
    =-g^{1/2}\,J^{\mu\nu}+\mathrm{O}[\,R_{\mu\nu}^3\,],     \label{GH.5.3}
   \end{eqnarray}
where $J^{\mu\nu}=J^{\mu\nu}[\,g\,]$ is the following rather
complicated nonlocal expression quadratic in Ricci tensor
   \begin{eqnarray}
     J^{\mu\nu}&=&\;g^{\mu\nu}\left\{\,
    -R^{\alpha\beta}\,\frac1\Box R_{\alpha\beta}
    +\Box\left[\,\frac14\Big(\frac1\Box R_{\alpha\beta}\Big)\,
    \frac1\Box R^{\alpha\beta}
    +\frac18\Big(\frac1\Box R\Big)^2\right]\right.\nonumber\\
    &&\qquad\qquad\qquad\quad\left.
    -2\left(\nabla^\gamma\frac1\Box R^{\alpha\beta}\right)
    \nabla_\alpha\frac1\Box R_{\beta\gamma}
    \right\}
    \nonumber\\
    &+&\;g^{\mu\nu}\frac1\Box
    \left\{\,\frac14 R^2-\frac12
    R_{\alpha\beta}^2+\frac14\Box R\,\frac1\Box R
    +\frac12\Box
    R^{\alpha\beta}\,\frac1\Box R_{\alpha\beta}
    -\nabla_\alpha\nabla_\beta R\,\frac1\Box R^{\alpha\beta}\right.
    \nonumber\\
    &&\qquad\qquad\qquad\quad\left.
    -R^{\alpha\beta}\,\nabla_\alpha\nabla_\beta \frac1\Box R
    +2\left(\nabla^\alpha\nabla^\beta\frac1\Box
    R^{\gamma\delta}\right)
    \nabla_\gamma\nabla_\delta \frac1\Box R_{\alpha\beta}\,\right\}
    \nonumber\\
    &-&\frac12\left(\nabla^\mu\frac1\Box R\right) \nabla^\nu\frac1\Box R
    -R^{\mu\nu}\,\frac1\Box R+\left(\nabla^{(\mu}\frac1\Box
    R^{\nu)\alpha}\right) \nabla_\alpha\frac1\Box R
    \nonumber\\
    &&\qquad
    -\left(\nabla^{(\mu}\nabla_\alpha \frac1\Box R\right)
    \frac1\Box R^{\nu)\alpha}
    -\left(\nabla^\mu\frac1\Box R_{\alpha\beta}\right)
    \nabla^\nu\frac1\Box R^{\alpha\beta}
    +2\left(\frac1\Box
    R^{\alpha\beta}\right)
    \nabla_\alpha\nabla_\beta\frac1\Box R^{\mu\nu}
    \nonumber\\
    &+&\nabla^{(\mu}\frac1\Box\left\{R^{\nu)\alpha}\,
    \nabla_\alpha\frac1\Box R+\nabla_\alpha R\,
    \frac1\Box R^{\nu)\alpha}+2 R_{\alpha\beta}\,\nabla^{\nu)}\frac1\Box
    R^{\alpha\beta}
    \right.\nonumber\\
    &&\qquad\qquad\left.+4\left(\nabla^{\nu)}\nabla^\alpha
    \frac1\Box R^{\beta\gamma}\right)
    \nabla_\gamma\frac1\Box R_{\alpha\beta}-4
    \left(\nabla^{\alpha}\nabla^\beta
    \frac1\Box R^{\nu)\gamma}\right)
    \nabla_\gamma\frac1\Box R_{\alpha\beta}\,
    \right\}.                                     \label{GH.5.4}
    \end{eqnarray}

Here we systematically integrated by parts neglecting the surface
terms which vanish for $\delta g_{\mu\nu}(x)$ with a compact
support and took into account that the commutator of covariant
derivatives contributes to the next order of the expansion in
Ricci curvatures. This commutator contains the Riemannian tensor
which in virtue of the Bianchi identities can be expressed as the
nonlocal power series in Ricci tensor \cite{CPTII}
   \begin{eqnarray}
    R^{\mu\nu\alpha\beta}=2\nabla^{[\mu}\nabla^\alpha\frac1\Box
    R^{\nu]\beta}-2\nabla^{[\mu}\nabla^\beta\frac1\Box
    R^{\nu]\alpha}
    +\mathrm{O}[\,R_{\mu\nu}^2\,].
   \end{eqnarray}
As a result the sum of three terms (\ref{GH.5.1})-(\ref{GH.5.3})
cancels in the quadratic approximation (corresponding to the cubic
approximation for $\Sigma[\,g\,]$), and the expression
(\ref{5.15}) turns out to be topologically invariant with respect
to local metric variations.

To obtain the dependence of $\Sigma=\Sigma[\,g_\infty\,]$ on the
asymptotic behavior of metric at infinity one should, instead of
the functional derivative, calculate its variation in the class of
$\delta g_{\mu\nu}(x)\sim 1/|x|^{d-2}$. Repeating the same steps
as above, one finds that only surface terms at infinity will
survive and, moreover, all surface integrals generated by
nonlinear terms of (\ref{5.15}) vanish in view of this falloff.
The only nonvanishing surface integral comes from the Einstein
part linear in the curvature and reads
   \begin{eqnarray}
    \delta_g\Sigma=\int
    dx\,g^{1/2}\,\left(\nabla^\mu\nabla^\nu\delta
    g_{\mu\nu}-g^{\alpha\beta}\Box\delta g_{\alpha\beta}\right)=
    \int_\infty d\sigma^\mu\;\delta^{\alpha\beta}
     (\partial_\alpha
     \delta g_{\beta\mu}-\partial_\mu \delta g_{\alpha\beta}),
   \end{eqnarray}
where the surface integration measure and the metric contracting
indices can be taken flat-space ones again in view of falloff
properties of $g_{\mu\nu}(x)$ and $\delta g_{\mu\nu}(x)$ at
$|x|\to\infty$. Thus, the right hand side can be easily
functionally integrated to give
(\ref{5.15})\footnote{Alternatively, Eq.(\ref{5.15}) can be
derived by expanding its left hand side in powers of metric
perturbations, $g_{\mu\nu}(x) =\delta_{\mu\nu}+h_{\mu\nu}(x)$, and
directly observing the cancellation of the bulk part, while the
nonvanishing surface terms reduce to the noncovariant form of the
Gibbons-Hawking surface integral.}.

\section{Metric dependence in the subleading order}
\hspace{\parindent} Using the expression (\ref{5.2}) for
$K(s\,|\,x,y)$ and the relation
$f^{\mu\nu}(\nabla_x,\nabla_y)\,\sigma(x,y)\,\big|_{y=x}=g^{\mu\nu}$
one has
    \begin{eqnarray}
    &&\frac{\delta{\Tr} K(s)}
    {\delta g_{\mu\nu}(x)}
    =-s\,g^{1/2}(x)\,f^{\mu\nu}(\nabla_x,\nabla_y)
    K(s|x,y)\,\Big|_{x=y}\nonumber\\
    &&\qquad\qquad=
    \frac{g^{1/2}(x)}{(4\pi s)^{d/2}}\,\left[
    \,\,\frac12 \,g^{\mu\nu}\,\Omega(s|\,x,x)
    -s\,f^{\mu\nu}(\nabla_x,\nabla_y)\,
    \Omega(s|\,x,y)\,\Big|_{\,y=x}\,\right],     \label{C.1}
    \end{eqnarray}
so that the metric variational derivative for $W_1$ looks more
complicated than (\ref{5.11a}) and has a contribution from the
leading order
    \begin{eqnarray}
    \frac{\delta W_1}{\delta g_{\mu\nu}}=
    \frac12\,g^{1/2}g^{\mu\nu}\Phi^2(x)-g^{1/2}
    f^{\mu\nu}(\nabla_x,\nabla_y)\,
    \Omega_1(x,y)\,\Big|_{\,y=x}\,.             \label{C.2}
    \end{eqnarray}

Checking this relation for $W_1$, given by Eq. (\ref{5.34}) with
metric-independent $W_1^\mathrm{metric}$, begins with the
calculation of $\Omega_1(x,y)$-contribution to the right hand side
of (\ref{C.2})
    \begin{eqnarray}
    &&-g^{1/2}f^{\mu\nu}(\nabla_x,\nabla_y)\,
    \Omega_1(x,y)\,\Big|_{\,y=x}=
    -\frac12\,g^{1/2}f^{\mu\nu}(\nabla_x,\nabla_y)\,
    \Big[\,\psi(x,y)+\psi(y,x)\,\Big]_{\,y=x}\nonumber\\
    &&\qquad\qquad\quad
    +\frac12\,g^{1/2}f^{\mu\nu}(\nabla_x,\nabla_y)\,
    \left[\,\frac1{F_x}\stackrel{\rightarrow}{F}_x\!
    \Phi(x)\,\sigma(x,y)\,\Phi(y)\!
    \stackrel{\leftarrow}{F}_y\!
    \frac{\stackrel{\leftarrow}{1}}{F_y}\;
    \right]_{\,y=x},                            \label{C.3}
    \end{eqnarray}
where $\psi(x,y)$ is given by Eq.(\ref{5.32}) and we use obvious
abbreviations $F_x=F(\nabla_x)$, etc. From the structure of the
operator $f^{\mu\nu}(\nabla_x,\nabla_y)$, (\ref{5.11}), it follows
that
    \begin{eqnarray}
    &&f^{\mu\nu}(\nabla_x,\nabla_y)\,
    \Big[\,\psi(x,y)-\psi(y,x)\,\Big]_{\,y=x}\nonumber\\
    &&\qquad\qquad\qquad\qquad\qquad
    =\frac12\,
    g^{\mu\nu}\,\Big[\,F_x\,\psi(x,y)-F_x\,
    \Psi(y,x)\,\Big]_{\,y=x},                   \label{C.4}
    \end{eqnarray}
and the two terms here can be simplified to
    \begin{eqnarray}
    &&F_x\,\psi(x,y)\,\Big|_{\,y=x}=\Phi(x)\,
    \left[\,\stackrel{\rightarrow}{F}_x\,\Phi(x)\,
    \sigma(x,y)\,\right]_{\,y=x}-d\,\Phi^2(x)=0,  \label{}\\
    &&F_x\,\psi(y,x)\,\Big|_{\,y=x}=\left.
    \frac1{F_x}\stackrel{\rightarrow}{F}_x[\,\Phi(x)\,
    \sigma(x,y)\,\Phi(y)\,]
    \stackrel{\leftarrow}{F}_y\,\right|_{\,y=x}.     \label{C.5}
    \end{eqnarray}
Therefore, the expression (\ref{C.2}) takes the form in which
$\psi(y,x)$ enters without symmetrization of its arguments
    \begin{eqnarray}
    &&-g^{1/2}\left.f^{\mu\nu}(\nabla_x,\nabla_y)\,
    \Omega_1(x,y){\vphantom{\frac11}}\,\right|_{\,y=x}=\left.
    \frac14\,g^{1/2}g^{\mu\nu}\,
    \frac1{F_x}
    \stackrel{\rightarrow}{F}_x[\,\Phi(x)\,\sigma(x,y)\,\Phi(y)\,]
    \stackrel{\leftarrow}{F}_y\,\right|_{\,y=x}
    \nonumber\\
    &&\qquad\qquad\qquad
    -\,g^{1/2}f^{\mu\nu}(\nabla_x,\nabla_y)\,\left.
    \psi(y,x){\vphantom{\frac11}}\,\right|_{\,y=x}
    \nonumber\\
    &&\qquad\qquad\qquad
    +\frac12\,g^{1/2}f^{\mu\nu}(\nabla_x,\nabla_y)\,
    \left[\frac1{F_x}
    \stackrel{\rightarrow}{F}_x \Phi(x)\,\sigma(x,y)\,\Phi(y)
    \stackrel{\leftarrow}{F}_y\!
    \frac{\stackrel{\leftarrow}{1}}{F_y}\;\right]_{\,y=x}.  \label{C.6}
    \end{eqnarray}

On the other hand, the metric variational derivative of the
nonlocal functional $W_1$, (\ref{5.34}), with the metric
independent $W_1^\mathrm{metric}$, (\ref{5.34a}), consists of the
following five terms
    \begin{eqnarray}
    &&\frac12\,\frac\delta{\delta g_{\mu\nu}(x)}
    \int dz\,g^{1/2}(z)\,\frac1{F_z}
    \stackrel{\rightarrow}{F}_z\,\Phi(z)\,\sigma(z,y)\,\Phi(y)
    \stackrel{\leftarrow}{F}_y\,\Big|_{\,y=z}\nonumber\\
    &&\qquad\quad\;=\frac12\,
    \int dz\,g^{1/2}(z)\,\frac1{F_z}
    \stackrel{\rightarrow}{F}_z\,\Phi(z)\,
    \frac{\delta\sigma(z,y)}{\delta g_{\mu\nu}(x)}\,\Phi(y)
    \stackrel{\leftarrow}{F}_y\,\Big|_{\,y=z}\nonumber\\
    &&\qquad\quad +
    \frac14\,g^{1/2}g^{\mu\nu}\,
    \frac1{F_x}
    \stackrel{\rightarrow}{F}_x \Phi(x)\,\sigma(x,y)\,\Phi(y)
    \stackrel{\leftarrow}{F}_y\,\Big|_{\,y=x}
    \nonumber\\
    &&\qquad\quad
    +\frac12\,g^{1/2}f^{\mu\nu}(\nabla_x,\nabla_y)\,
    \Big(\frac1{F_x}
    \stackrel{\rightarrow}{F}_x \Phi(x)\,\sigma(x,y)\,\Phi(y)
    \stackrel{\leftarrow}{F}_y\!
    \frac{\stackrel{\leftarrow}{1}}{F_y}\;\Big)_{\,y=x}\nonumber\\
    &&\qquad\quad
    -\,g^{1/2}f^{\mu\nu}(\nabla_x,\nabla_y)\,
    \Big(\,\Phi(x)\,\sigma(x,y)\,\Phi(y)\,]
    \stackrel{\leftarrow}{F}_y\!
    \frac{\stackrel{\leftarrow}{1}}{F_y}\;\Big)_{\,y=x}\nonumber\\
    &&\qquad\quad
    +\int dz\,g^{1/2}(z)\,\frac1{F_z}
    \stackrel{\rightarrow}{F}_z\,G(z,x)\,
    f^{\mu\nu}(\stackrel{\rightarrow}{\nabla}_x,
    \stackrel{\leftarrow}{\nabla}_x)\,
    \Phi(x)\,\sigma(z,y)\,\Phi(y)\!
    \stackrel{\leftarrow}{F}_y\Big|_{\,y=z}.   \label{C.7}
    \end{eqnarray}
The first three terms correspond to the variation of the world
function, measure $g^{1/2}(z)$ and Green's function $1/F_z$
respectively, whereas the last two ones -- to the variation of
operators ($F_z$, $F_y$) and the functions $(\Phi(z),\Phi(y))$.

In the first term one can integrate by parts so that the operator
$F_z$ will act to the left and "cancel" the Green's function. The
corresponding surface term will be zero for the following reason.
The variational derivative $\delta\sigma(z,y)/\delta
g_{\mu\nu}(x)$ is not vanishing at infinity, but on the sphere of
infinitely growing radius $|z|\to\infty$ it has a support only at
the point where the geodesic emanating from the point $y$ and
passing through $x$ punctures this sphere. The contribution of
this point is suppressed to zero by the angular measure
$\sim\delta(\theta)\,\sin^{d-1}\theta$, where $\theta$ is the
longitudinal spherical angle (with the origin at the north pole
$\theta=0$ coinciding with this point). Integration by parts then
yields the local expression
    \begin{eqnarray}
    &&\frac12\,
    \int dz\,g^{1/2}(z)\,\frac1{F_z}
    \stackrel{\rightarrow}{F}_z\,\Phi(z)\!\left.
    \frac{\delta\sigma(z,y)}{\delta g_{\mu\nu}(x)}\,\Phi(y)\!
    \stackrel{\leftarrow}{F}_y\right|_{\,y=z}\nonumber\\
    &&\qquad\quad=\frac12\,
    \int dz\,g^{1/2}(z)\,\Phi(z)\,F(\nabla_y)\left.
    \frac{\delta\sigma(z,y)}{\delta g_{\mu\nu}(x)}\,\Phi(y)
    \,\right|_{\,y=z}=
    \frac12\,g^{1/2}g^{\mu\nu}\,\Phi^2(x),       \label{C.8}
    \end{eqnarray}
based on simple coincidence limits
    \begin{eqnarray}
    &&\left.\nabla_\alpha^y\,\frac{\delta\sigma(y,z)}{\delta g_{\mu\nu}(x)}
    \,\right|_{\,y=z}=0,  \nonumber\\
    &&\left.\Box_y\,\frac{\delta\sigma(y,z)}{\delta g_{\mu\nu}(x)}
    \,\right|_{\,y=z}=\frac{\delta}{\delta
    g_{\mu\nu}}\Big(\Box_y\sigma(y,z)\,|_{z=y}\Big)-
    \left.\frac{\delta\,\Box_y}
    {\delta g_{\mu\nu}}\,\sigma(y,z)
    \,\right|_{\,y=z}=g^{\mu\nu}\delta(z,x).       \label{C.9}
    \end{eqnarray}
This expression obviously reproduces the first term on the right
hand side of Eq.(\ref{C.2}).

In the last term of (\ref{C.7}) integration by parts over $z$ is
again possible without extra surface terms because
$G(z,x)\sigma(z,y)\,\Phi(y)\!\stackrel{\leftarrow}{F}_y\sim1/|z|^{d-3}$,
therefore the $(z,y)$-part of the integrand reduces to the local
coincidence limit
$\sigma(z,y)\,\Phi(y)\!\stackrel{\leftarrow}{F}_y\big|_{\,y=z}=d\,\Phi(z)$
and the result looks
    \begin{eqnarray}
    &&\int dz\,g^{1/2}(z)\,\frac1{F_z}
    \stackrel{\rightarrow}{F}_z\,G(z,x)\,
    f^{\mu\nu}(\stackrel{\rightarrow}{\nabla}_x,
    \stackrel{\leftarrow}{\nabla}_x)\,
    \Phi(x)\,\sigma(z,y)\,\Phi(y)\!
    \stackrel{\leftarrow}{F}_y\Big|_{\,y=z}\nonumber\\
    &&\qquad\qquad\qquad\qquad\qquad =g^{1/2}
    f^{\mu\nu}(\nabla_x,\nabla_y)\,\left[\,d\,\Phi(x)\,
    \Phi(y)\frac{\stackrel{\leftarrow}{1}}{F_y}
    \,\right]_{\,y=x}.                          \label{C.10}
    \end{eqnarray}
The fourth term of (\ref{C.7}) together with (\ref{C.10}) equals
$-f^{\mu\nu}(\nabla_x,\nabla_y)\, \psi(y,x)\,\big|_{\,y=x}$. Thus
the last four terms of (\ref{C.7}) reproduce the contribution of
$\Omega_1(x,y)$, Eq.(\ref{C.6}), to the metric variational
derivative (\ref{C.2}). Taken together with (\ref{C.8}) this
completes the proof of the equation (\ref{C.2}) for the subleading
coefficient $W_1$ given by (\ref{5.34}) with a constant
(metric-independent) $W_1^\mathrm{metric}$.

\end{document}